\documentclass[%
 reprint,
nofootinbib,
 amsmath,amssymb,
 aps,
 pra,
]{revtex4-2}

\usepackage{graphicx}
\usepackage{dcolumn}
\usepackage{bm}
\usepackage{hyperref}
\usepackage{mathtools}
\usepackage[capitalise]{cleveref}

\usepackage[colorinlistoftodos,prependcaption,textsize=footnotesize]{todonotes}

\DeclarePairedDelimiter\norm{\lVert}{\rVert}

\usepackage[most]{tcolorbox}
\tcbuselibrary{listings,breakable}

\definecolor{helleresblau}{HTML}{3B67CE}
\definecolor{codegreen}{HTML}{33998d}
\definecolor{codegray}{rgb}{0.5,0.5,0.5}
\definecolor{lilablau}{HTML}{100094}

\hypersetup{
	colorlinks=true,
	linkcolor=lilablau,
	citecolor=lilablau,
	urlcolor=lilablau,
	final=true,
	plainpages=false,
	pdfstartview=FitV,
	pdftoolbar=true,
	pdfmenubar=true,
	pdfencoding=auto,
	psdextra,
	bookmarksopen=true,
	bookmarksnumbered=true,
	pdfnewwindow=true,
	breaklinks=true,
	linktocpage=true
}

\newtcblisting{docstring}[1][]{
    breakable,
    listing only,
    colback=white,
    colframe=codegreen,
    boxrule=0.5pt,
    arc=0mm,
    title={#1},
    fonttitle=\bfseries,
    listing options={
        basicstyle=\ttfamily\footnotesize,
        breaklines=true,
        columns=fullflexible,
        literate={é}{{\'e}}1
         {·}{{$\cdot$}}1
         {≈}{{$\approx$}}1
         {→}{{$\rightarrow$}}1
         {×}{{$\times$}}1
         {Δ}{{$\Delta$}}1
         {Γ}{{$\Gamma$}}1
         {≥}{{$\geq$}}1
         {±}{{$\pm$}}1
         {–}{{\textendash}}1
         {∈}{{$\in$}}1
         {−}{{$-$}}1
         {°}{{$\degree$}}1
         {~}{{$\sim$}}1,
    }
}

\usepackage{gensymb}

\newtcblisting{prompt}[1][]{
    breakable,
    listing only,
    colback=white,
    colframe=helleresblau,
    boxrule=0.5pt,
    arc=0mm,
    title={#1},
    fonttitle=\bfseries,
    listing options={
        basicstyle=\ttfamily\footnotesize,
        breaklines=true,
        columns=fullflexible,
        literate={é}{{\'e}}1
         {·}{{$\cdot$}}1
         {≈}{{$\approx$}}1
         {→}{{$\rightarrow$}}1
         {×}{{$\times$}}1
         {Δ}{{$\Delta$}}1
         {Γ}{{$\Gamma$}}1
         {≥}{{$\geq$}}1
         {±}{{$\pm$}}1
         {–}{{\textendash}}1
         {∈}{{$\in$}}1
         {−}{{$-$}}1
         {°}{{$\degree$}}1
         {~}{{$\sim$}}1,
    }
}

\newtcblisting{pycode}[1][]{
    breakable,
    listing only,
    colback=white,
    colframe=codegreen,
    boxrule=0.5pt,
    arc=0mm,
    title={#1},
    fonttitle=\bfseries,
    left=8mm,
    listing options={
        language=Python,
        basicstyle=\ttfamily\footnotesize,
        keywordstyle=\color{codegreen}\bfseries,
        stringstyle=\color{helleresblau},
        commentstyle=\color{codegray}\itshape,
        showstringspaces=false,
        breaklines=true,
        columns=fixed,
        keepspaces=true,
        tabsize=4,
        upquote=true,
        numbers=left,
        numberstyle=\tiny\color{codegray},
        literate={é}{{\'e}}1
         {·}{{$\cdot$}}1
         {≈}{{$\approx$}}1
         {→}{{$\rightarrow$}}1
         {×}{{$\times$}}1
         {Δ}{{$\Delta$}}1
         {Γ}{{$\Gamma$}}1
         {≥}{{$\geq$}}1
         {±}{{$\pm$}}1
         {–}{{\textendash}}1
         {∈}{{$\in$}}1
         {−}{{$-$}}1
         {°}{{$\degree$}}1
         {~}{{$\sim$}}1,
    }
}

\DeclareMathOperator{\grad}{grad}

\usepackage{xcolor}
\usepackage{soul}
\sethlcolor{yellow}

\begin{document}

\title{Fully Automated Discovery of Phase Diagrams Using Agentic Large Language Models}

\author{Robert Amelung}
\email{robert.amelung@fau.de}

\author{Maximilian Nägele}
\email{maximilian.naegele@mpl.mpg.de}

\author{Florian Marquardt}

\affiliation{Max Planck Institute for the Science of Light, Staudtstraße 2, 91058 Erlangen, Germany\\Department of Physics, Friedrich-Alexander-Universität Erlangen-Nürnberg, Staudtstraße 5, 91058 Erlangen, Germany}

\date{\today}

\newcommand{\pch}{\textsc{Phasonaut}}
\newcommand{\sci}{\textsc{sciexplorer}}

\begin{abstract}

Classifying phases of matter and mapping out phase diagrams are central problems in physics. Recently, machine learning techniques have emerged as promising tools for phase classification, ranging from supervised learning to various unsupervised methods. However, these techniques provide limited qualitative insight into the discovered phases,  typically work only for specific system classes, and do not by themselves efficiently map out the phase diagram.  Autonomously constructing the phase diagram of a general system purely from measurements, without prior knowledge of the system, while also yielding qualitative understanding of the phases, remains an open task. In this work, we introduce \pch, a two-agent framework that leverages the tool-use capabilities and broad knowledge of large language models (LLMs) to drive an active learning loop that constructs phase diagrams of physical systems and provides interpretations of the discovered phases. Starting from scratch, it iteratively selects and performs measurements, analyzes the resulting data, and segments the parameter domain into the phases it discovers. \pch\ recovers the phase diagrams of classical and quantum spin systems to substantial precision. In the future, this framework could enable rapid exploration of systems with unknown phase diagrams, including in autonomously driven experiments.

\end{abstract}

\maketitle

\section{Introduction}

Throughout physics, determining the different phases of a system is a central challenge. Phase diagrams provide a qualitative understanding of a system’s behavior by compressing its many possible configurations into a small number of regimes with similar properties. In doing so, they reveal how microscopic interactions give rise to emergent macroscopic phenomena, and which microscopic details are relevant for them. They can also guide the search for materials with desired properties and applications. However, identifying phases typically requires both broad physical intuition and system-specific considerations, for example to choose suitable order parameters. As a result, mapping phase diagrams remains largely a manual process.

Over the past decade, machine-learning approaches have emerged as a promising route toward automatically distinguishing phases~\cite{dawidModernApplicationsMachine2025}. When phase-labeled data is available, neural networks are able to recognize phases from previously unseen simulated~\cite{carrasquillaMachineLearningPhases2017,sunDeepLearningTopological2018,zhangMachineLearningTopological2018, PhysRevX.7.031038,dongMachineLearningQuantum2019} or experimental~\cite{remIdentifyingQuantumPhase2019, linkMachineLearningPhase2023} data after supervised training. However, often no such labeled data is available, and unsupervised approaches have to be employed. These include principal component analyses~\cite{wangDiscoveringPhaseTransitions2016,wetzelUnsupervisedLearningPhase2017,huDiscoveringPhasesPhase2017}, clustering and manifold learning~\cite{Rodriguez2019, Chng2018}, autoencoders~\cite{huDiscoveringPhasesPhase2017,wetzelUnsupervisedLearningPhase2017}, anomaly detection~\cite{kottmannUnsupervisedPhaseDiscovery2020,kottmannUnsupervisedMappingPhase2021,kamingUnsupervisedMachineLearning2021}, confusion-based methods~\cite{vannieuwenburgLearningPhaseTransitions2017,vannieuwenburgLearningPhaseTransitions2018,bohrdtAnalyzingNonequilibriumQuantum2021,Bruder2024}, and network-prediction analyses~\cite{schaferVectorFieldDivergence2019,greplovaUnsupervisedIdentificationTopological2020,hoRobustIdentificationTopological2021,hoSelfsupervisedEnsembleLearning2023}. These methods have successfully been applied to both conventional symmetry-breaking transitions associated with local order parameters, and topological ones where order is inherently non-local. However, each of these tools typically performs well only on certain system classes. Moreover, they do not provide qualitative understanding of the phases, i.e., they do not yield order parameters associated with the transitions on their own. Different approaches try to tackle this problem~\cite{ponteKernelMethodsInterpretable2017,greitemannProbingHiddenSpin2019,sadouneLearningSymmetryprotectedTopological2026,SymmetriesMaciej2021}, e.g., by analyzing weights or activations of a trained neural network~\cite{wetzelMachineLearningExplicit2017, Miles2021Correlator,cybiński2026tetriscnninterpretabledetectionphases}. Human input and analysis are still crucial for operating these methods and understanding their results.

To automatically find different phases present in a system without labeled data, the search can be steered by gradient descent through a differentiable ground-state solver~\cite{zhou2026autonomous}, but this requires a differentiable simulation. To instead locate entire phase boundaries, confusion-based methods can be combined with segmentation algorithms from computer vision~\cite{liuDiscriminativeCooperativeNetworks2018}. However, this can only separate two distinct regions, instead of autonomously mapping out the full diagram, and relies on manually choosing reasonable initial conditions. Often, gathering data is the most expensive part of mapping out a phase diagram. If experimental data can be phase-labeled, active learning methods lower the cost by adaptively choosing the most informative measurements. For example, Bayesian active learning efficiently locates phase boundaries~\cite{GaussianProcessesForActiveLearningOfPDsPaper1, GaussianProcessesForActiveLearningOfPDsPaper2, mcdannald2022}. However, these techniques either ignore physical reasoning or rely on physical constraints hand-coded for specific systems in advance.

\begin{figure*}
    \centering
    \includegraphics[width=\textwidth]{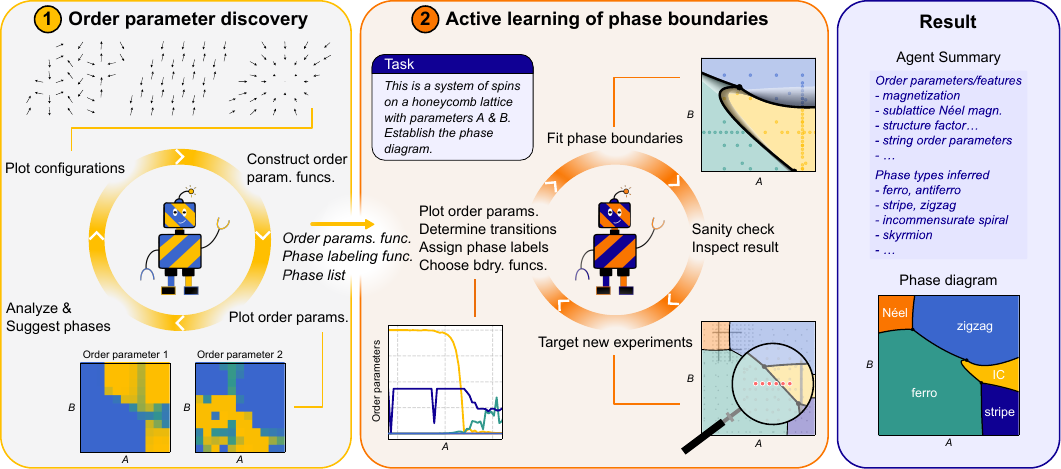}
    \caption{Agentic setup of \pch. The phase discovery agent (left) analyzes experiments, computes and plots order parameters or raw experimental data, and finally proposes a list of phases present in the parameter domain, together with coded utilities for computing the relevant parameters. This is represented here for the case of classical ground-state spin lattice experiments, where the phase discovery agent studies spin configurations over a coarse experimental sweep of the parameter domain. The results are automatically passed to the active learning agent (center), whose task is to construct the full phase diagram through a process of iterative targeting of experiments and fitting of phase boundaries. It analyzes data -- using the phase discovery agent's results as a baseline --, assigns phase labels to experiments, and uses the segmentation tool to fit self-defined parametric curves to said labeling for the phase boundaries. When satisfied, the active learning agent provides a final phase diagram in standardized form (see \cref{ssec:SegTool}), together with plain-text explanations of the phases' features. Both agents are asked to summarize their explorations.}
    \label{fig:AgenticSetup}
\end{figure*}

Recently, large language model (LLM) agents have shown promise in automating tasks that are still difficult for traditional machine-learning methods, because they draw upon prior knowledge, and can adapt to previously unknown situations. Pioneering works in chemistry show that they can plan and execute experiments in automated laboratories~\cite{boikoAutonomousChemicalResearch2023, chemcrow}, and in mathematics, they autonomously prove theorems~\cite{polu2020generative}. In physics, LLM agents control superconducting qubit experiments~\cite{CAO2025101372,li2026large}, atomic force microscopes~\cite{mandal2025}, and particle accelerators~\cite{Hellert2026}, run simulations~\cite{gustinAgenteCuanticoAutomating2026}, design photonic integrated circuits~\cite{Poon2025}, propose research ideas in quantum optics and turn them into concrete experimental designs~\cite{arlt2025autonomousquantumphysicsresearch}, and recover physical models from data~\cite{Sciexplorer,kleidermacherDiscoveringModelsExperimental2026}. Recently, LLMs were incorporated into a specific part of a phase-diagram discovery process by deciding which experiments to perform next while mapping the phase diagram of an alloy system~\cite{tamuraLLMguidedPhaseDiagram2026}. However, the phases themselves were still identified by human experts.

To date, no fully autonomous framework exists for determining the phase diagram of general systems purely from measurements that also provides physical insight about the discovered phases. To address these limitations, we introduce \pch, a two-agent framework that leverages the agentic capabilities and broad physical knowledge of modern LLMs~\cite{gridachAgenticAIScientific2025, zhu2025probing} to explore and map out unknown phase diagrams in a fully autonomous approach. \pch\ automatically runs an active learning loop in a given black-box system to target experiments, analyze the results with self-written code, visualize data, and iteratively refine its predicted phase boundaries.

We show that \pch\ recovers a selection of phase diagrams of classical and quantum spin lattice models with substantial accuracy, and further discuss the application of our framework to a yet unexplored model. We demonstrate that boosting \pch's performance is possible by leveraging its separation into two agents, as well as by making use of self-criticism abilities of modern LLMs~\cite{madaan2023selfrefine, Sciexplorer}. Additionally, we show that the active learning strategies employed by \pch\ surpass naive grid searches. This work opens the way towards fully autonomous phase-diagram explorations of unknown systems, including programmatically controlled real experiments.

\section{Agentic \pch}
\label{sec:AgenticPch}

An LLM-based agent is a language model running in a loop that can call external tools. It reads its current context and prompts, reasons internally, and decides which tool to call next. It then reads the results and repeats until the task is completed. \pch\ is an LLM-based \hbox{two-agent} framework that autonomously discovers two-dimensional phase diagrams of physical systems. It autonomously performs experiments, analyzes data, and segments a fixed parameter domain into the phases it discovers (see \cref{fig:AgenticSetup}). To save and later retrieve large tool results that do not fit the LLM's context window, \pch\ uses an external memory. The ``experiments" are numerical simulations which, from the point of view of the agents, originate from a black-box system. \pch\ builds on the \sci\ framework, a recently proposed LLM-based agent designed for autonomous discovery of physical models~\cite{Sciexplorer}.

This LLM-agent-based approach has some fundamental advantages compared to other phase-classification methods. Foremost, prior physical knowledge enters directly through the pretraining of the LLM. On one hand, this includes knowledge about different areas of physics (e.g.\ statistical physics), and more specifically, potential phases and order parameters, the influence of geometry and dimensionality, and types of phase transitions. These can be particularly helpful in determining distinct phases present in an unknown system. On the other hand, LLMs are also aware of finite-size effects and scaling, and typical features of phase diagrams, which guide the localization of precise phase boundaries. In contrast, typical unsupervised phase-classification methods utilize this knowledge at most indirectly, e.g., via specific choices to represent the experimental data, symmetries directly encoded in neural-network architectures, or finite-size scaling in post-processing. Moreover, these methods follow a fixed procedure: they cannot tailor their approach to the specific system type under study, nor adjust their analysis in response to intermediate results, while LLM agents do both through reasoning.

\subsection{Setup}

Exploring the phase diagram of a physical system can typically be separated into an initial ``analysis" stage that determines the distinct phases present within the parameter domain, and a subsequent ``active learning" stage to discover the precise phase-boundary locations via a suitable sequence of measurements. Motivated by this general approach, \pch\ consists of two separate agents, which we refer to as ``Phase Discovery Agent" (PDA) and ``Active Learning Agent" (ALA). Advantages of this separation include better focus for each agent, since their respective tasks are narrower, and smaller context windows for the LLMs. The modularization further enables boosting performance by running several PDAs in parallel (\cref{ssec:MultiplePDAs}).

\pch\ can perform experiments by specifying points in the two-dimensional parameter domain. Parameters could represent, for example, a coupling constant or an external field, but this information is unknown to the agents. An experimental result can be, for instance, a ground-state spin configuration in a classical spin system. As a first step, \pch\ gathers initial experimental data by measuring over a coarse grid in the parameter domain. The PDA analyzes this data by defining suitable order parameters (OPs), e.g., magnetizations, structure factors and chirality measures. When it is finished, it returns its analysis in a standardized form: both a fixed set of OPs which it deems relevant for distinguishing phases in the given system, and a function that maps sets of these OPs to proposed phase labels. In order to facilitate interpretability by humans or fact-checking by the ALA, it also supplies explanations for the chosen OPs, the function, and the discovered phases. We use the term ``order parameter" in a broad sense, encompassing all types of features and observables that characterize a given state, including those associated with unconventional or exotic order.

These results are fed to the ALA, whose task is to perform further targeted experiments in order to chart the two-dimensional phase diagram with high precision. In principle, it can also stress-test and improve the PDA's analysis based on the additional data. When satisfied, the ALA provides the final result in standardized form. It supplies the full phase diagram in the fixed parameter domain as a set of phase boundaries, and labels the thus segmented regions explicitly. Since the specific shape of different boundaries can be of interest, these are directly provided as parametric curves defined in Python. For interpretability, the ALA also provides a final list of the discovered phases -- associated with regions in the diagram -- with explanations.

For human inspection or checking by separate agents, both the PDA and the ALA are asked to summarize their explorations after they have provided results.

Through a process of trial and error, we developed separate system prompts for both agents that guide the general analysis and active learning loops. The goal being a capable autonomous phase explorer, as opposed to a general-purpose artificial scientist, we provide detailed ``to-do lists" including also general information about different types of phases. As a first step, the PDA plots experimental data and candidate OPs to form an initial assessment of the possible phases. Through iterative code-based and visual analysis, it then updates its hypothesis for which phases are present and which OPs are relevant. In our view, this general loop would also be performed by a human researcher. The ALA, on the other hand, performs an iterative phase-boundary refinement, also typical of human procedure. To produce an initial ansatz for the phase diagram, it assigns phase labels to experiments and determines approximate boundary crossings. It then defines parametric curves in parameter space for the phase boundaries, which are automatically fitted to the labeled experiments. To further refine the diagram, it targets experiments in the most uncertain regions. These steps are not strictly enforced, however, and the agents interpret and implement the given rules differently in different attempts, according to what they deem most suitable. 

During their respective explorations, both agents have access to sets of tools and descriptions thereof. To make use of modern LLM coding and vision capabilities, we do not provide specialized analysis or visualization tools to the agents. Instead, both agents can execute self-written Python code, create images, and save intermediate results in an external memory.

Fitting phase boundary curves to a given phase labeling of a set of points in parameter space is a reoccurring task for the ALA that involves a collection of technical choices. Currently available LLMs still display difficulties related to this task. Therefore, we provide a ``segmentation tool" to the ALA, which we discuss in the next section.

Furthermore, \pch\ can use the tools that give access to experiments. All tool results are saved in the external memory and can subsequently be analyzed by the agent. We limit the experiment budget since we want to avoid fine-grid searches during active learning.

In this work, we discuss results of applying \pch\ to classical and quantum spin lattice systems at zero temperature. This choice is driven by simplicity, and the entire agentic setup would also work for finite temperatures without changes to the agents. Every model is parametrized by two real parameters \(A\) and \(B\). In the first prompt, the agents are informed about their task and the parameter domain on which they should explore the diagram. Minimal information related to the type of system and lattice (e.g.\ ``classical continuous spins on a 2D triangular lattice with \(N\) sites in both dimensions") is also provided. As a rule, information of this general form would also be available in real experiments even in cases where the specific underlying model is unknown. They are further informed about technical details such as conventions related to lattice encodings or system-size and sweep limitations. They are asked explicitly to infer the thermodynamic limit (infinite size) phase diagram from finite-size experiments. Importantly, neither the Hamiltonian nor any further model-specific information is provided; \pch\ explores phase diagrams solely from experimental data, as would be necessary in a real experiment.

In the classical setup, an experiment yields the full ground-state spin configuration for a parameter point \((A,B)\) and for the system size that the agent provides. The results of the PDA are, formally, (i)~a Python function mapping spin configurations to tuples of relevant OPs, (ii)~a Python function mapping such tuples to proposed phase labels, (iii)~descriptions of both functions, and (iv)~a plain-text list of the discovered phases. This setup is shown in \cref{fig:AgenticSetup}.

In the quantum setup, an experiment is the measurement of the ground-state expectation value of an operator, for a parameter point \((A,B)\) and a system size that the agent provides. For efficiency, several operators can be measured in a single tool call. The operators are defined by the agent using a separate tool. We explicitly do not return the quantum ground state directly, since real quantum mechanical measurements do not provide it. This differs fundamentally from the classical setup, where we provide the full ground state. Instead of a function computing OPs from spin configurations, the PDA returns a list of operator definitions whose expectation values are relevant and a Python function mapping tuples of such expectation values to proposed phase labels, with descriptions.

\subsection{Segmentation tool}
\label{ssec:SegTool}

The most naive approach to mapping out a (two-dimensional) phase diagram is to assign phase labels to points on a regular grid in the parameter domain. However, we aim for explicit boundary curves, since they define exact separations of phases that are not limited by some grid resolution, and because they offer better interpretability. We find that currently available LLM models still display difficulties related to consistent fitting of phase boundaries to experimental data, when tasked to explicitly define such boundary curves on their own: boundaries often do not meet precisely at their end-points, fitting techniques are inconsistent and results have higher spread as a consequence (see, for instance, \cref{fig:Ablations}\,(c) in the appendix). For this reason, we provide the ALA with a segmentation tool that automatically fits given boundary shapes to labeled points in parameter space. To use the tool, the ALA is compelled to reason more diligently about the global structure of the diagram and the functional form of different boundaries.

To make use of the tool, the agent must fix the structure of the suspected diagram. It provides the adjacency structure of the multicritical points (MCPs) as a directed graph, including the possibility of phase boundaries starting or ending at the domain borders. We make use of the LLM's intuition to obtain general boundary shapes, but finding precise parameters usually needs computation. For this reason, the agent provides Python code defining parametric curves
\begin{align}
\begin{split}
    f_\text{bdry}: [0,1]\times \mathbb{R}^k&\rightarrow \mathbb{R}^2,\\
    (t, \mathbf{p}) & \mapsto (A(t, \mathbf{p}), B(t, \mathbf{p}))
\end{split}
\end{align}
for every phase boundary. Each boundary can have a distinct shape and parameters \(\mathbf{p}\) of varying dimension~\(k\). The agent provides phase-labeled points \((A,B)\) which are to be segmented, optionally complemented by points expected to lie on the boundaries, and an initial guess for the parameters \(\mathbf{p}\) and MCP positions.

\begin{figure}
    \centering
    \includegraphics[width=\columnwidth]{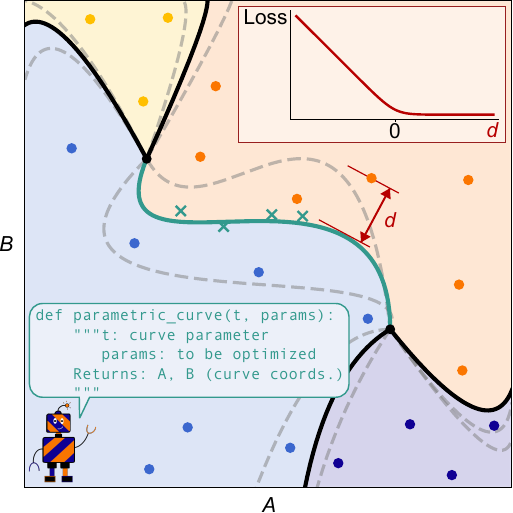}
    \caption{Segmentation tool. The tool fits a given phase-boundary structure to phase-labeled points in the parameter domain, which are shown as colored dots. The active learning agent provides Python code defining parametric curves for each boundary, which can depend on additional parameters. Dashed lines suggest different segmentations associated with different values of these parameters. Boundary points can optionally be provided, represented here with green crosses. Each labeled point contributes to the loss function depending on their signed distance \(d\) to the closest boundary. The optimization aims at finding the best possible segmentation of the labeled points, given the functional constraints and the adjacency structure of the multicritical points.}
    \label{fig:SegTool}
\end{figure}

Internally, the tool performs gradient-based minimization of a loss function reflecting the quality of a given segmentation, by varying the parameters \(\mathbf{p}\) and the MCP positions. Each labeled point contributes a loss dependent on its position relative to the closest boundary. If it lies on the correct side of the boundary, we assign a positive distance \(d\) to the point. Conversely, the assigned distance is negative if the side is incorrect. The loss contribution is given by a scaled softplus function
\begin{equation}
    \mathcal{L}(d) \propto \varepsilon \ln (1+\mathrm{e}^{-d/\varepsilon}),
\end{equation}
which vanishes quickly when the point is correctly segmented. In addition, a soft constraint is imposed on the end points of each boundary, so that they coincide at the MCPs or lie on the domain borders. A number of further soft constraints aims to ensure the absence of intersections between boundaries, and the containment of all boundaries and MCPs within the domain. For more detail, we refer the reader to \cref{app:SegToolAppendix}. A schematic overview of the segmentation tool is shown in \cref{fig:SegTool}.

\begin{figure*}
    \centering
    \includegraphics[width=\textwidth]{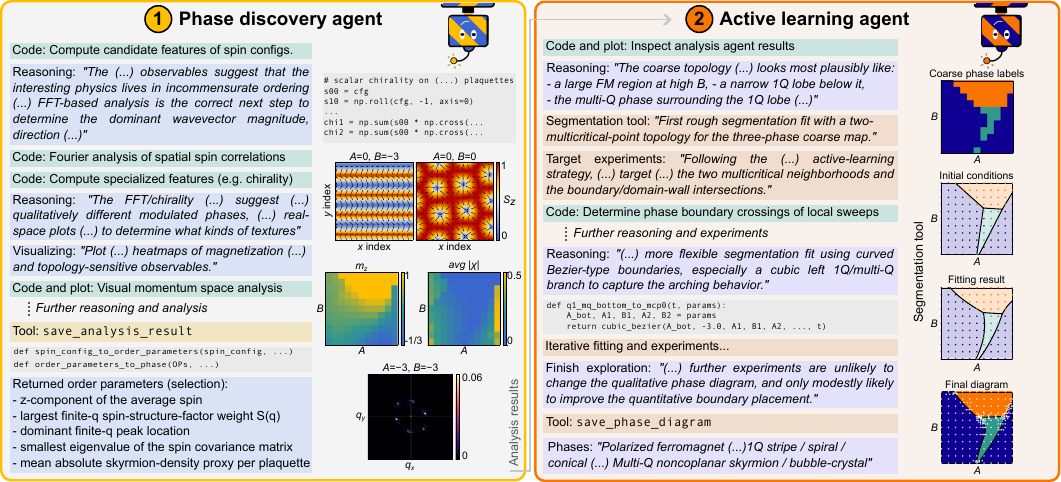}
    \caption{Example phase-diagram exploration by \pch. All plots, code, and quotes shown in this figure were created by the agents; colors have been adapted. The explored model is an extension of a classical model known to exhibit skyrmionic crystals~\cite{SkrymionBaselineModelForNewModel} by an additional chiral plaquette term. The phase discovery agent (left) computes and plots features of spin configurations in a sweep, including real-space visualization of configurations (upper plots) and feature heatmaps (center plots: \(z\)-magnetization and averaged scalar chirality). It computes specialized features and performs Fourier analysis of the spatial spin correlations (lower plot: momentum space of a skyrmionic configuration). A selection of the order parameters returned by the phase discovery agent's \texttt{spin\_config\_to\_order\_parameters} is shown. The active learning agent (right) first inspects the analysis results, plotting phase labels on a coarse grid. It then iteratively performs experiments and uses the segmentation tool to increase the precision of all phase boundary locations. The final predicted diagram consists of a \(z\)-polarized region (orange), a single-\(\mathbf{q}\) spiral phase (green), and a skyrmionic crystal phase (blue). The locations of all experiments performed are overlaid (white dots). We used GPT-5.4 with a budget of 1{,}000 experiments, although only 241 were used.}
    \label{fig:Exploration}
\end{figure*}

To help the agent in its further process, we supply secondary information in addition to the fitted parameters and MCP positions, including the fraction of correctly segmented points\footnote{Importantly, without reference to the ground truth. The accuracy returned here refers only to the labeling provided by the agent.} to judge the fit quality, and signed distances \(d\) for each point to spot potentially wrongly labeled outliers. To directly provide a visual overview, the result also contains colored images of the phase diagram described by both the initial guess and the best fit. Additionally, the agent automatically gets informed when it likely provided a contradictory structure. In that case, the listed secondary information can also help the agent correct its mistakes.

The input format for the segmentation tool is also used for the final result tool, which the ALA calls at the end of its exploration. Consequently, the ALA can simply input the best fit that the segmentation tool provided in its opinion.

\section{Results}

\pch\ determines coherent and well-founded phase diagrams in most cases. We discuss specific aspects of the performance in the following sections. For some models, there can be significant spread in the results, and we discuss methods to address this, e.g., by comparing multiple runs using a separate LLM critic or combining multiple PDA results (\cref{ssec:Benchmarking,,ssec:MultiplePDAs}). An example exploration is shown in \cref{fig:Exploration}.

In a typical exploration, the PDA starts by computing (or measuring) a range of diagnostic observables or features on a coarse grid in the parameter domain. For classical systems, these include magnetizations, spin-component statistics, scalar chirality, planarity metrics and the spin structure factor. For quantum systems, typically analyzed features include magnetizations (e.g.\ squared, Néel-staggered, different spin axes), correlators (e.g.\ nearest-neighbor), dimer operators, chiral operators, parity operators, and string operators. The PDA usually plots pixelated heatmaps of varying selections of these quantities in order to visually discern contiguous regions with similar properties. For classical systems, it also inspects real-space plots of the spin configurations. After gaining initial understanding from these diagnostics, the agent investigates specific features more closely that it finds relevant, such as peak distributions in the structure factor. It executes code to guide its working hypothesis for which types of phases are present and inspects some regions more closely when they do not fit the general picture, or to decide whether it should split a phase into sub-phases. A selection of common phases and OPs is mentioned in the PDA's system prompt (see supplementary material~\cite{SM}). Crucially, while the prompts differ between the classical and quantum setup, it is kept the same for all models of each setup. Moreover, the PDA goes beyond the provided examples as needed, considering also more specialized phases (e.g.\ skyrmions) and OPs (e.g.\ chirality measures). When the agent concludes its analysis, it usually tests the functions it must provide before settling on the final result, checking that all phases found are represented.

The ALA usually starts by assessing the PDA's results. In the classical setup, for instance, it can immediately plot a pixelated heatmap of the proposed phases using the initial sweep. After deciding whether the PDA's results are sound~-- possibly by executing additional code~-- it begins to determine the precise boundaries. Using plotted coarse phase maps, the ALA first decides on a rough topology of the diagram (e.g.\ ``polarized phase in the upper central region, \dots"), which it sometimes directly inputs as a base fit to the segmentation tool. Subsequently, it targets relevant locations to perform experiments, usually near suspected boundaries or multicritical points. Often, this is complemented by finite-size scaling analysis, e.g., Binder cumulant crossings in the quantum setup. The choice of boundary shapes generally varies, but linear and higher-order Bézier curves are used in most cases. When satisfied, the best result from a segmentation tool call is usually provided as the final result.

On average, the exploration of a classical model takes about one hour, with a standard deviation of about fifteen minutes. The PDA accounts for about fifteen minutes of that runtime. The exploration of a quantum model takes longer due the long simulation times for large system sizes: about \(1.8\) hours total with a standard deviation of about \(40\) minutes, for which the PDA accounts about \(20\) minutes. A fairer comparison is the number of steps used: Both classical and quantum explorations use on average about \(40\) steps for the ALA (\(\sigma\approx 8\)) and \(24\) steps for the PDA (\(\sigma\approx 5.5\)). These statistics correspond to the benchmarks discussed in \cref{ssec:Benchmarking}.

\subsection{Application to an unexplored model}
\label{ssec:Exploration}

In general, models whose phase diagrams are known have likely contributed to the training of current LLMs. However, \pch\ can directly be applied to systems whose phase diagrams have not been explored so far, since it does not rely on prior knowledge of the system.

To demonstrate its abilities, we ran \pch\ on a classical continuous-spin lattice model which, to our knowledge, has not been studied in this form. Through independent numerical analysis using the same experiment tool, we confirm that the final predicted diagram and the phases identified seem plausible. Naturally, human verification and further analysis remain a necessary step in assessing the outcome of any autonomous AI-based scientific exploration.

Excerpts from the exploration with GPT-5.4~\cite{openai_gpt54} are shown in \cref{fig:Exploration}, including the final predicted diagram. We used an extension of a model known to exhibit skyrmionic crystals~\cite{SkrymionBaselineModelForNewModel} by an additional chiral plaquette term, on the two-dimensional square lattice with periodic boundary conditions (see \cref{app:ModelsAppendix}). The parameters scale the chirality term (\(A\)) and an external magnetic field (\(B\)). In this exploration, the PDA quickly understands that skyrmionic crystals are involved, and the spiral phase is determined using adequate Fourier analysis of spatial spin correlations (structure factor). The multi-\(\mathbf{q}\) ordering of the skyrmions is visualized directly in momentum space and compared to the other phases. Given the PDA's results, the ALA decides that the phase diagram likely consists of three separate phases. Phase boundaries are iteratively refined with segmentation tool calls and a total of 241 experiments including the initial sweep.

\subsection{Benchmarking}
\label{ssec:Benchmarking}

In typical scientific applications, LLM agents are used for problems where candidate answers can automatically be verified, e.g., mathematical proofs verified in Lean~\cite{ren2025deepseekproverv2advancingformalmathematical} or model discovery tasks in physics~\cite{Sciexplorer}, where a proposed model can be empirically checked by comparing its predictions to measurements. Phase-diagram discovery differs from such tasks, since a proposed solution cannot easily be checked, assuming we do not know the true phase diagram beforehand. To still gain some insight into \pch's performance, we benchmark a select number of classical and quantum spin lattice systems of which the ground-truth zero-temperature phase diagrams are known. One might argue that this is, consequently, not a representative study, since the LLMs could have learned the ground-truth diagrams during training. However, we usually do not observe any modeling behavior or retrieval from ``memory" in these benchmarked runs. We provide further discussion of these aspects near the end of this section. These benchmarks are run with \mbox{OpenAI's} GPT-5.2~\cite{openai_gpt52} on ``high" reasoning setting for both agents.

We benchmark \pch\ on both classical and quantum systems. Specifically, we consider classical spin lattice models with continuous spins. To simulate them, we use gradient-based energy minimization starting from random initial spin configurations (in a parallelized batch) for increased accuracy (see \cref{app:SimulationDetailsAppendix}). We set a budget of 400 experiments for the agent and limit the number of lattice sites to \(100\) for one-dimensional models, and to \(60^2\) for two-dimensional models due to compute constraints. The considered quantum models are also spin lattice systems, and we use exact diagonalization of the Hamiltonian to compute expectation values in the ground state (see \cref{app:SimulationDetailsAppendix}). We set a limit of \(4{,}000\) experiments for both agents combined. The ten-fold increase compared to the classical benchmarks is motivated by the fact that a quantum measurement reveals only partial information about the system's state. Any remaining budget from the PDA is added to the ALA's budget. Due to compute time, we limit quantum benchmarks to one-dimensional chains.

To assess the accuracy of an individual run's resulting diagram, we employ an averaged Intersection-over-Union (IoU) measure known from object detection and other applications~\cite{IoUpaper1,IoUpaper2,IoUpaper3}. This weights all ground-truth phases equally, irrespective of their size, and significantly punishes the absence of a true phase in the predicted diagram. Concretely, we compute the averaged normalized overlap between phases in the prediction and the ground truth:
\begin{equation}
\label{eq:AccuracyMeasure}
    \text{Accuracy} = \frac{1}{M} \sum_{i=1}^M \frac{A_{i,\text{pred}}\cap A_{i,\text{gt}}}{A_{i,\text{pred}}\cup A_{i,\text{gt}}},
\end{equation}
where \(M\) is the number of distinct phases in the ground truth, and \(A_{i,\text{pred/gt}}\) are the regions (areas) of the parameter domain belonging to phase \(i\) in prediction and ground truth, respectively. We approximate the accuracy numerically by sampling a fine grid of the domain.

\begin{figure}[h!]
    \centering
    \includegraphics[width=\columnwidth]{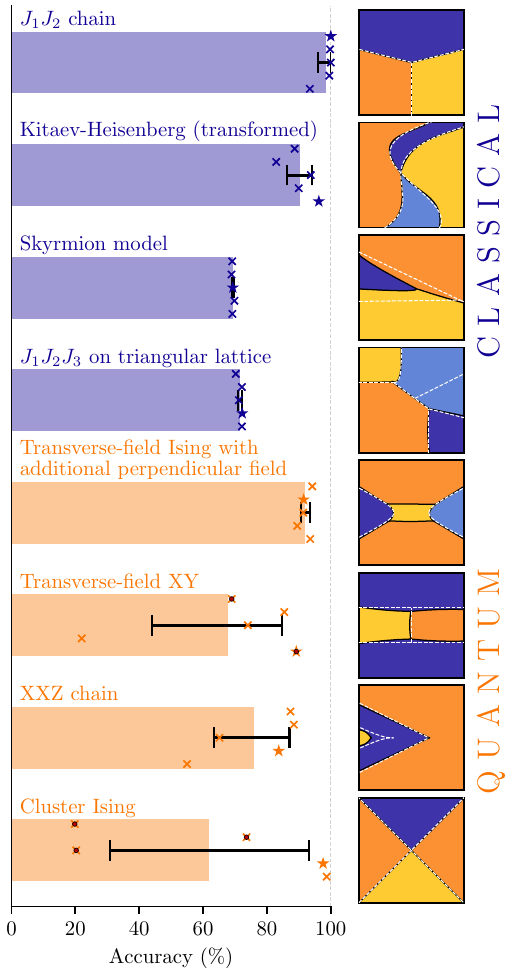}
    \caption{Benchmarks for \pch. We test four classical (blue) and four quantum (orange) spin lattice models with known phase diagrams; see \cref{app:ModelsAppendix} for the models' Hamiltonians. Five runs are performed for each model, the result accuracies of which (see \cref{eq:AccuracyMeasure}) are shown as crosses and stars. Bars denote the mean accuracy for each model, and bootstrapped \(95\,\%\) confidence intervals for the means are shown. On the right, the respective results of highest accuracy are shown; the ground-truth phase boundaries are overlaid with white dashed lines. Stars denote the ``best runs" as designated by the LLM critic when asked to review the runs without knowing the ground truth. Red dots denote runs where the agent's qualitative description of the phases was inaccurate.}
    \label{fig:Benchmarks}
\end{figure}

We benchmark four classical and four quantum models, the results of which are shown in \cref{fig:Benchmarks}. We run each model five times and compute bootstrapped \(95\,\%\) confidence intervals for the mean accuracy. For reference, we list the precise Hamiltonians used in \cref{app:ModelsAppendix} and only give short descriptions here.

For the classical benchmarks, we run a nearest-neighbor and next-nearest-neighbor Heisenberg chain (\(J_1J_2\) chain in \cref{fig:Benchmarks}), a parameter-transformed Kitaev-Heisenberg model on the honeycomb lattice, a specific model known to display a skyrmion crystal phase on the triangular lattice~\cite{SkyrmionBenchmarkModelRef} (``Skyrmion model"), and a planar \(J_1J_2J_3\) model on the triangular lattice (\(J_1J_2J_3\)-triangle in the following). The phases represented include (anti)ferromagnetism, polarization along an external field, spin spiral waves, specific configurations on the honeycomb lattice and skyrmions. The results for these models display generally high accuracy. In the \(J_1J_2J_3\)-triangle model, one of the phases needed to be split into two qualitatively distinct sub-phases, which the agents did in none of the runs, explaining their lower accuracy scores. Accuracies in the Skyrmion model may be biased due to the imperfect nature of our numerical simulation method (see \cref{app:SimulationDetailsAppendix}), but the results show that the agents are capable of discovering exotic phases.

In the quantum setup, we benchmark the transverse-field Ising model with additional perpendicular field (TFIMwPF in the following), the transverse-field XY model, the XXZ chain, and the cluster Ising model. The first three models display conventional ferromagnetic, antiferromagnetic and polarized or paramagnetic phases, associated with different spin orientations. In our definition, the cluster Ising model consists of a polarized and a symmetry-protected topological phase. Generally, the conventional phases are discovered correctly, but segmentations vary between runs. The latter observation seems to be more frequent in the quantum than in the classical explorations, which we attribute to it generally being more difficult to decide on precise boundaries due to finite-size effects. Sometimes, additional confusion arose due to degenerate points (or lines) in the parameter space, where we chose to return expectation values with respect to a mixed ground state. The agents sometimes decided to devote spatially extended phases to those degeneracies. We note that the central (yellow) phase in TFIMwPF found in each run does not actually exist and should be merged with the upper and lower (orange) phase. This central phase is usually designated as ``\(y\)-polarized" by the agent, in contrast to ``\(x\)-polarized" in the upper and lower regions. This is not entirely far-fetched, since only the constant \(y\)-component of the external field remains near \(B=0\), but there is no phase transition separating this supposed central phase from the upper and lower regions. For some of the quantum explorations, the final qualitative description of at least one phase was not entirely correct. In \cref{fig:Benchmarks}, these runs are marked with an additional red dot. In particular, the topological phase in the cluster Ising model was found only twice out of five attempts. For the transverse-field XY model, ``polarized" (along an external field) and ``ferromagnetic" were exchanged in these cases.

One method to boost \pch's performance beyond that of a single run is to leverage the significant ability of modern LLMs to conduct independent critical reviews of LLM-generated output. To this end, we let fresh GPT-5.2 instances review the complete conversations and the final diagram of all runs of a given model, and decide which of the five attempts best solved the scientific problem. Importantly, these critics do not have access to the corresponding accuracy values nor the underlying model or ground-truth phase diagram. To overcome LLM context-window limitations, we first let a separate review be created for each run; these reviews were then shown to a separate ``decider" LLM. We ran the latter five times for each model, randomizing the order in which the reviews were supplied, and settled on the run chosen most often (for the prompts used, see the supplemental material~\cite{SM}). For all but two models, the decider chose the same run in all five attempts. The thus designated ``best runs" are marked in \cref{fig:Benchmarks} as stars. In general, this technique boosts performance despite the fact that phase-diagram discovery constitutes a particularly challenging scientific task where the result cannot be verified without access to some externally supplied ground truth (in contrast to, e.g., model discovery, where predictions and observations can be compared).

One could argue that \pch\ does well on these benchmarks because it recognizes the models and knows their phase diagrams. However, we find that even though the agents sometimes guess the correct model, or model type, in their reasoning traces, they usually do not seem to simply reproduce diagrams from ``memory". Instead, they perform accurate analyses of the experiments and segment the diagram accordingly. This can be seen, for instance, in the Kitaev-Heisenberg model, where we pre-transformed the parameter space in order to make the phase boundaries non-linear: the agents do not simply copy the linear boundaries of the original Kitaev-Heisenberg model, instead they actively learn the precise boundaries. Furthermore, the model explored in \cref{ssec:Exploration} does not have, to our knowledge, a well-known ground-truth phase diagram; nevertheless, the agents propose one, using their usual analysis techniques. This suggests that \pch\ yields sensible results also for models whose ground-truth diagrams are unknown.

\subsection{Multiple phase discovery agents}
\label{ssec:MultiplePDAs}

Another way to improve the framework's performance leverages the separation of \pch\ into two agents. It enables running multiple PDAs in parallel, which all return their results (order parameters and phase labeling functions) to the ALA. In the best case, the latter combines the results and decides in particular which PDAs obtained the most accurate results. As a consequence, the non-verifiable setting can be boosted simply with additional analysis. This approach has complementary advantages to using an LLM critic to compare multiple full runs (as discussed above): re-running only the PDA has a lower computational cost compared to running the full pipeline. Crucially, if all PDAs base their analysis on the same experimental data (as is the case here), this is even possible without performing any additional (typically costly) experiments.

\begin{figure}
    \centering
    \includegraphics[width=\columnwidth]{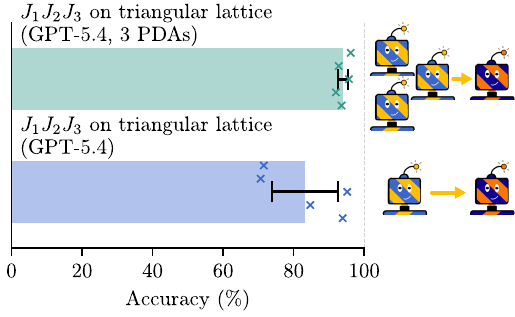}
    \caption{Boosted performance using multiple phase discovery agents. Results of five runs with three phase discovery agents (3 PDAs, green) and five runs with a single phase discovery agent (light blue), for the \(J_1J_2J_3\)-triangle classical model. Single-run accuracies are shown as crosses, bars denote the respective mean accuracies, and bootstrapped \(95\,\%\) confidence intervals for the means are shown.}
    \label{fig:3PDA-Comparison}
\end{figure}

We ran a three-PDA setup with GPT-5.4 for \(J_1J_2J_3\)-triangle, feeding all three analysis results to a single ALA. In a slightly modified system prompt, the ALA was told to combine the results and choose the union of all proposed phases that proved robust. To obtain statistics, we repeated this entire setup five times, and, for comparison, we did the same for the standard single-PDA setup. The results are shown in \cref{fig:3PDA-Comparison}, displaying a significant increase in accuracy for the three-PDA runs.

In this model, the method proves advantageous since the upper right incommensurate-ordered phase in \(J_1J_2J_3\)-triangle is not consistently split by a PDA run~-- although it should be, since the two sub-phases correspond to two qualitatively distinct incommensurate spiral-wave patterns on the triangular lattice. In all five three-PDA runs, at least one PDA split the phase correctly, and the ALA decided to incorporate it into the final diagram, leading to high accuracies. In the single-PDA runs, only three out of five runs split the phase, leading to a significant decrease in accuracy compared to the three-PDA setup. Note that for these runs we chose GPT-5.4, which displays higher rates of splitting the phase than GPT-5.2.

\begin{figure*}
    \centering
    \includegraphics[width=\textwidth]{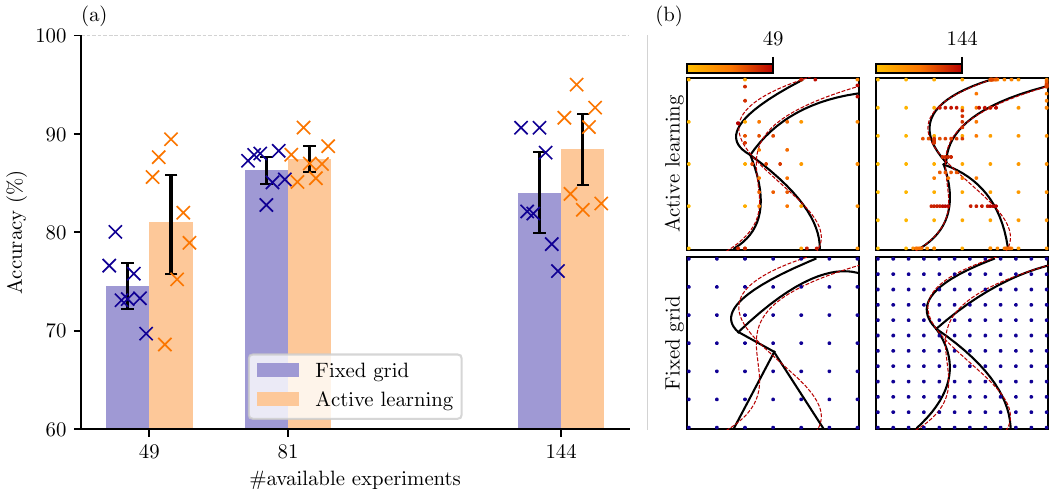}
    \caption{Assessing the active learning of \pch. (a) Results for the active learning capabilities benchmarking on the classical transformed Kitaev-Heisenberg model. For the same experiment budgets of \(49\), \(81\) and \(144\), we compare standard ``Active learning" runs (orange) -- where the agent can target experiments -- with ``Fixed grid" runs (blue) where the agent merely has access to a predefined regular grid in the parameter domain. In both cases, we did not run any phase discovery agent; instead, a previously obtained high-quality result was returned to every active learning agent (see supplemental material~\cite{SM}). Each setup was run seven times with GPT-5.4. The accuracies are shown as crosses; bars indicate mean accuracies for each setup, and bootstrapped \(95\,\%\) confidence intervals for the means are shown. Despite significant spread in all categories, we observe an overall advantage of agentic active learning strategies over naive grid search. (b) Highest-accuracy results for \(49\) (left) and \(144\) (right) available experiments, with overlaid experiment locations (active learning: top, fixed grid: bottom). The color scale in the active learning runs indicates the order in which the experiments were performed. The predicted phase boundaries are shown in black, while the ground truth is overlaid with red dashed lines. In the active learning runs, the agent did not always use its entire budget.}
    \label{fig:ALvsNAL}
\end{figure*}

\subsection{Active learning capabilitites}
\label{ssec:ALCapabilities}

Through their reasoning, LLMs are able to drive an active learning loop, using heuristics to adaptively target experiments. This reasoning is performed by the ALA in \pch. In its system prompt, we encourage the ALA to target experiments where the boundaries are still most uncertain, and to reason about functional forms for the boundaries. Furthermore, it can improve its understanding of the physical system's phases from iterative experimentation.

To evaluate the active learning component of \pch, it is desirable to minimize the variability of other components, particularly the analysis performed by the PDA. To this end, we bypass the PDA by always returning identical high-quality analysis results to the ALA which we obtained from a standard run during testing.

As a model system, we consider the classical transformed Kitaev-Heisenberg model since it displays non-trivial boundary shapes. We compare performances between ALAs that are allowed to target experiments as usual, and ALAs that are handed a fixed grid sweep of experiments at the beginning of their exploration and thus cannot use active learning for targeting experiments. One may argue that the number of experiment results given to a PDA would generally influence the quality of that PDA's results, but we neglect this here and simply focus on assessing the active learning aspect given a reasonable fixed analysis. We explored three different choices of experiment budgets, of \(49\), \(81\) and \(144\), and each of those was run seven times to obtain statistics. The results are shown in \cref{fig:ALvsNAL}. Although there is some statistical spread, partly attributable to the ALA's variable use of the segmentation tool, we find that on average, active learning strategies employed by the agent lead to higher accuracies compared to the fixed-grid runs. In panel (b), the highest-accuracy runs' results for an experiment budget of \(49\) and \(144\) are shown with overlaid experiment locations. In the active learning runs, one can observe linear sweeps that cross suspected boundaries, as well as single experiments near more uncertain segments. This example shows that agentic active learning can improve accuracy while retaining potential advantages of on-the-go physical reasoning.

Broadly, we assess that GPT-5.4 generally seems to adhere to instructions in its system prompt more closely than GPT-5.2. We thus decided to use GPT-5.4 for this comparison. With GPT-5.2, we were unable to show a significant advantage of targeted experiments over fixed grid experiments.

\subsection{Failure modes and limitations}

LLMs are known to hallucinate, leading to false reasoning or invention of facts. These general limitations associated with agentic systems seem to be, however, at least partly mitigated in phase-diagram explorations due to the iterative self-correcting and self-review necessary for the task. Nevertheless, \pch\ tends to fixate on its hypotheses when it believes to have discovered the general picture of the diagram, rarely back-tracking entirely when new data contradicts its chain of thought. This stubbornness is also displayed in the ALA's use of the segmentation tool. Boundaries are sometimes kept linear out of principle, even if the fit suggests a more flexible form is needed; conversely, too complex boundary forms are sometimes tried as a first step, never going back to linear boundaries to test whether they would suffice for an accurate description of the boundary.

A further common failure mode are low-quality PDA results. These include, for instance: proposed phase-label functions that do not map to the entirety of discovered phases given the default threshold values set by the PDA; missed phases, or incorrectly splitting a well-defined phase; over-focusing on degenerate points or experiment artifacts, sometimes defining separate phases for such findings. Low-quality analysis usually translates into final low-accuracy results, since the ALA typically does not strongly question the analysis results.

The ALA sometimes does not successfully correct contradictions in its structural input to the segmentation tool, such as inverted parametrizations of a boundary or mixing up the left- and right-side phase labeling of single boundaries; in that case, results are nonsensical but can typically be sorted out by a human researcher checking the diagram. Additionally, the agent sometimes acts too confidently, e.g., ignoring the instruction to make use of the segmentation tool in favor of implementing a (worse) custom fit. Furthermore, complex phase structures are often difficult for the ALA to fit, sometimes leading to nonsensical results when the segmentation tool is used improperly.

While some of these challenges, such as missed phases, can partially be solved by the methods introduced above (running multiple PDAs; using an LLM critic to choose the best attempt out of many), others are more difficult to solve, likely requiring more capable LLMs which might become available in the future.

\subsection{Ablation experiments}
\label{ssec:Ablations}

We performed ablation experiments on the quantum model TFIMwPF and the classical model \(J_1J_2J_3\)-triangle in order to assess the effect of different components of \pch\ on its final accuracy. The results are shown in \cref{fig:Ablations}. Removing the PDA and transferring its task to the ALA has negligible effect on the accuracy in both models. Availability of the segmentation tool does not seem to boost our accuracy measure (\cref{eq:AccuracyMeasure}); however, it leads to more consistent phase-diagram structures with smooth, connected boundaries. Replacing all system prompts by default non-specific instructions reduces performance in the quantum model, but only slightly in the classical model. See \cref{app:Ablations} for a detailed discussion.

\section{Discussion and conclusion}

In this work, we introduced an LLM-based two-agent framework that autonomously explores and maps out the phase diagram of black-box physical systems, purely from simulated experimental data. Through iterative analysis with coding and visualizations, the phase discovery agent determines order parameters, and proposes directly interpretable phases for the diagram. Using these results, the active learning agent iteratively targets experiments to determine the precise phase boundaries, using the additional experimental knowledge gained to further refine its understanding of the different phases and their transitions.

\pch\ re-discovers known phase diagrams of classical and quantum spin lattice models to substantial precision, but also explores unknown systems, including both conventional and topological systems. We studied the most prominent failure modes occurring in this challenging task, such as missing a phase due to overconfidence. Detecting such failures is complicated by the fact that the quality of a proposed phase diagram cannot be determined automatically, i.e., phase-diagram discovery belongs to a class of very challenging tasks without straightforward ground-truth-independent verification. To address this, we propose methods to boost the framework's performance by using multiple phase discovery agents or leveraging the self-criticism capabilities of modern LLMs to review and compare repeated attempts. Moreover, \pch\ employs active learning strategies that are advantageous compared to grid search.

In practice, there is no guarantee for the correctness of the discovered diagram. To verify its quality, a human researcher can, for instance, compare repeated runs or task the agent to make predictions based on the diagram and check them experimentally. For example, the agent could be asked to describe the behavior of specific observables in a user-defined cut, possibly at a larger system size than was available during the exploration.

\pch\ explores phase diagrams solely based on observables without knowledge of the underlying model. In the future, it could therefore be directly applied to autonomously driven experiments, such as quantum simulators realized, e.g., by cold atomic gases or superconducting qubits.

During the final preparation of this manuscript, a preprint using an LLM agent to determine phase diagrams of quantum systems was released. As opposed to our manuscript, which considers working with an unknown system, that work focuses on a setting where the LLM agent is aware of the system's Hamiltonian and therefore does not choose experiments acting on a ``black-box" system, but instead simulates the systems using DMRG. This provides a complementary perspective and we refer the reader to their work~\cite{quantummatterExplorer}.

\section*{Data and code availability}

The code for the \pch\ framework, as well as logs of all runs, will be made public on~\cite{r-amelung-github} upon publication of this manuscript in a peer-reviewed journal.

\bibliography{references}

\setcounter{equation}{0}
\setcounter{figure}{0}
\setcounter{table}{0}
\setcounter{section}{0}

\renewcommand{\theequation}{A\arabic{equation}}
\renewcommand{\thefigure}{A\arabic{figure}}
\renewcommand{\thetable}{A\arabic{table}}
\renewcommand{\thesection}{A\arabic{section}}

\renewcommand{\theHequation}{A\arabic{equation}}
\renewcommand{\theHfigure}{A\arabic{figure}}
\renewcommand{\theHtable}{A\arabic{table}}
\renewcommand{\theHsection}{A\arabic{section}}

\appendix
\onecolumngrid

\section{Models}
\label{app:ModelsAppendix}

In the following, we list the Hamiltonians of the models explored by \pch. For each model, the two variable parameters are \(A\) and \(B\), and the target domain is set to \((A,B)\in[-3,3]^2\).\\

\textit{Classical benchmark models.} The classical models assume normalized three-component spin vectors \(\vec{S}_i\) at every lattice site \(i\).

The \(J_1J_2\) chain model consists of nearest- and next-nearest-neighbor Heisenberg interactions on a one-dimensional chain with open boundary conditions:
\begin{equation}
    H_{J_1J_2} = A \sum_{\langle i,j\rangle} \vec{S}_i \cdot \vec{S}_j + B \sum_{\langle\langle i,j\rangle\rangle} \vec{S}_i \cdot \vec{S}_j.
\end{equation}
Nearest neighbors and next-nearest neighbors are denoted by \(\langle i,j\rangle\) and \(\langle\langle i,j\rangle\rangle\), respectively, where each pair is counted once. The phases separate into antiferromagnetic (\(A>0\)) and ferromagnetic (\(A<0\)) ordering when the next-nearest neighbor interaction does not exceed \(B=|A|/4\). Above these critical lines, helical spirals form.

The classical Kitaev-Heisenberg model is defined on the two-dimensional honeycomb lattice as~\cite{KitaevHeisenbergModelRef}
\begin{equation}
    H_\text{KitHeis} = \sum_{\gamma\in \{x,y,z\}} \sum_{\langle i,j \rangle_\gamma} J \vec{S}_i \cdot \vec{S}_j + K S_i^\gamma S_j^\gamma,
\end{equation}
where \(\gamma\) denotes the three different types of nearest-neighbor bonds. The phase boundaries are given by the linear relations \(K=-J\) and \(J=0\), which separate antiferromagnetic, ferromagnetic, ``zigzag" and ``stripy" configurations on the honeycomb lattice~\cite{KitaevHeisenbergModelRef}. To test \pch\ on complicated boundary shapes, we artificially curved these boundaries by reparametrizing \(J\) and \(K\) as
\begin{equation}
    J = A - \frac{B^2}{5} - \frac{B^3}{10} + \frac{3}{5},\qquad
    K = B - \frac{3}{25}.
\end{equation}
We imposed open boundary conditions for the transformed Kitaev-Heisenberg model.

The Skyrmion model benchmark was studied in~\cite{SkyrmionBenchmarkModelRef}; it is defined on the two-dimensional triangular lattice as
\begin{equation}
    H_\text{Skyrmion} = -\sum_{\langle i,j\rangle} \vec{S}_i\cdot\vec{S}_j + \frac{1}{2} \sum_{\langle\langle i,j\rangle\rangle} \vec{S}_i \cdot \vec{S}_j
    - \frac{K}{2} \sum_i (S_i^z)^2 - h\sum_i S_i^z.
\end{equation}
We used the phase diagram of~\cite[Figure 1]{SkyrmionBenchmarkModelRef}, obtained via numerical methods, as our ground truth, which includes a polarized phase, a spiral phase, and the central skyrmionic crystal phase. However, we restricted it to the domain \(K\in[0.25, 0.6]\), \(h\in[0.01, 0.3]\). This choice was due to our solver not consistently reproducing the other more subtle phases that appear outside this domain. The associated parameter transformation used is the following:
\begin{equation}
    K = \frac{7}{120} A + \frac{17}{40},\qquad
    h = \frac{29}{600}B + \frac{31}{200}.
\end{equation}
We imposed periodic boundary conditions because it improved the solver's performance for this model. 

Finally, the \(J_1J_2J_3\)-triangle classical benchmark model is an effectively planar nearest-, next-nearest, and third-nearest-neighbor model on the two-dimensional triangular lattice with open boundary conditions:
\begin{equation}
    H_{J_1J_2J_3} = -\sum_{\langle i,j\rangle} \vec{S}_i\cdot\vec{S}_j + A \sum_{\langle\langle i,j\rangle\rangle} \vec{S}_i \cdot \vec{S}_j
    + B \sum_{\langle\langle\langle i,j\rangle\rangle\rangle} \vec{S}_i \cdot \vec{S}_j + \sum_i (S_i^z)^2.
\end{equation}
Third-nearest neighbors are denoted by \(\langle\langle\langle i,j\rangle\rangle\rangle\), where each pair is counted once. The ground truth for this model is discussed in~\cite{NonSimpleMagneticOrderForSimpleHamiltonians}. All its phases can be described by single-\(\vec{q}\) ordering (spiral waves with wave vector \(\vec{q}\)). The model displays a ferromagnetic phase (\(\vec{q}=0\), lower left), commensurate-ordered spiral phases (\(\vec{q}=M,K\) in the Brillouin zone; lower right and upper left, respectively), and two incommensurate-ordered spiral phases, distinguished by the direction of the wave vector (\(\vec{q}\parallel \Gamma M\) or \(\vec{q}\parallel \Gamma K\); upper right, with the \(\vec{q}\parallel \Gamma M\) phase adjacent to the \(M\) phase, and the \(\vec{q}\parallel \Gamma K\) phase adjacent to the \(K\) phase). All boundaries are linear, with the exception of the \(K/\Gamma K\) boundary, which admits no closed form.\\

\textit{Unexplored model.} The model used for the example exploration in \cref{ssec:Exploration} is an extension of a model known to exhibit skyrmionic crystals~\cite{SkrymionBaselineModelForNewModel} by an additional chiral term, on the two-dimensional square lattice with periodic boundary conditions. The chiral term is a triple product over sets of three spins sitting on plaquettes of the square lattice. Such contributions arise in effective descriptions of the Fermi-Hubbard model in an external magnetic field in the large-\(U\) regime~\cite{LargeULimitHubbardChiralSpinInteractions}. The Hamiltonian used is given by
\begin{align}
\begin{split}
    H_\text{\cref{ssec:Exploration}} = -J  \sum_{i,\mu} \vec{S}_i \cdot \vec{S}_{i+\hat{\mu}}
        -A_c \sum_i ( S_i^y S_{i+\hat{x}}^y + S_i^x S_{i+\hat{y}}^x )
        -D  \sum_i \left[ \hat{y}\cdot (\vec{S}_i \times \vec{S}_{i+\hat{x}}) - \hat{x} \cdot (\vec{S}_i \times \vec{S}_{i+\hat{y}}) \right]\\
        +A_s \sum_i (S_i^z)^2 - h \sum_i S_i^z
        - \kappa \sum_{[ijkl]} \left(  \vec{S}_i \cdot (\vec{S}_j \times \vec{S}_k) + \vec{S}_i \cdot (\vec{S}_k \times \vec{S}_l \right),
\end{split}
\end{align}
where \([ijkl]\) denotes square plaquettes, and \(\hat{\mu}=\hat{x},\hat{y}\). We set \(J=1\), \(D=0.5\), \(A_c=D^2/2|J|=0.125\). We also set \(A_c + A_s = D^2/J = 0.25\), where the model in~\cite{SkrymionBaselineModelForNewModel} displays a spiral, skyrmionic, and ferromagnetic phase. The two variables are the chirality parameter \(\kappa\) and the external field \(h\), which were parametrized such that \(\kappa/J\in[-1,1]\) and \(hJ/D^2\in[0,3]\):
\begin{alignat}{2}
    \kappa/J &= \Tilde{A}, &\qquad h &= \Tilde{B} D^2/J,\\
    \Tilde{A} &= A/3, &\qquad \Tilde{B} &= (B+3)/2.
\end{alignat}

\textit{Quantum benchmark models.} The quantum models benchmarked in \cref{ssec:Benchmarking} are one-dimensional spin-\(1/2\) chains with open boundary conditions. Due to limited compute time, the number of lattice sites was capped at 12. The Hamiltonians read
\begin{equation}
\label{eq:TFIMwPF}
    \mathcal{H}_\text{TFIMwPF} = A\sum_{\langle i,j\rangle} \sigma_i^z \sigma_j^z - \sum_i \Big(2B\sigma_i^x + \sigma_i^y\Big),
\end{equation}
\begin{equation}
    \mathcal{H}_\text{XY} = -\frac{1}{2} \sum_{\langle i,j\rangle} \Big(\frac{1+A}{2}\sigma_i^x \sigma_j^x + \frac{1-A}{2}\sigma_i^y \sigma_j^y\Big)
    - \frac{B}{2} \sum_i \sigma_i^z,
\end{equation}
\begin{equation}
    \mathcal{H}_\text{XXZ} = -\frac{1}{4} \sum_{\langle i,j\rangle} \Big(\sigma_i^x \sigma_j^x + \sigma_i^y \sigma_j^y + A \sigma_i^z \sigma_j^z\Big)
     - B \sum_i \sigma_i^z,
\end{equation}
\begin{equation}
    \mathcal{H}_\text{CLIsing} = A\sum_{\langle i,j,k\rangle} \sigma_i^x \sigma_j^z \sigma_k^x + B \sum_i \sigma_i^z,
\end{equation}
respectively for the TFIMwPF, transverse-field XY, XXZ, and cluster Ising models. Standard Pauli matrices are denoted by \(\sigma_i^x, \sigma_i^y,\sigma_i^z\) for lattice site \(i\). Nearest neighbors are denoted by \(\langle i,j\rangle\), and a succession of three sites is denoted by \(\langle i,j,k\rangle\).

For TFIMwPF, note that the constant external field component \(\sigma^y\) is transverse with respect to the (anti)ferromagnetic alignment. The model is thus equivalent to a reparametrization of the transverse-field Ising model. The critical lines separating the polarized/paramagnetic phase from the ferro- and antiferromagnetic phases are hyperbolas described by \(A = \pm \sqrt{4B^2+1}\).

The transverse-field XY model and the XXZ chain are exactly integrable models~\cite{franchiniIntroductionIntegrableTechniques2017}. The former consists of \(x\)- and \(y\)-ferromagnetic phases (right and left central regions, respectively), separated by the \(A=0\) line, which compete against the polarized phase (upper and lower regions) that dominates for \(|B|>1\) in the thermodynamic limit.

The XXZ chain admits a polarized phase aligned with the \(z\)-ferromagnetic interaction if \(|B|>|(1-A)/2|\) or \(A>1\). A \(z\)-antiferromagnetic phase exists for \(A<-1\) and small external field \(B\). The critical lines separating this phase from an intermediate, paramagnetic phase, admit an implicit definition using elliptical integrals~\cite{franchiniIntroductionIntegrableTechniques2017}.

The cluster Ising model has a well-known phase diagram and is commonly used as a testing ground for machine learning approaches~\cite{sadouneLearningSymmetryprotectedTopological2026}. In our definition, we suppress the ferromagnetic contribution. The resulting diagram includes the polarized phase (upper and lower region), as well as the symmetry-protected topological phase (left and right regions), where the boundaries are given by \(B=\pm A\).

\section{Simulation details}
\label{app:SimulationDetailsAppendix}

In the classical setup, the Hamiltonian \(H\) is a function defined on the product manifold \((S^2)^k\) of normalized 3D spin vectors, where \(k\) is the number of sites. To take this into consideration, we perform Riemannian gradient descent with iterative renormalization: in each step, every spin vector gets rotated slightly according to
\begin{equation}
    \vec{S}_i\rightarrow \frac{\vec{S}_i-\eta\grad_i H}{\norm{\vec{S}_i-\eta\grad_i H}},
\end{equation}
where the gradient is projected onto the tangent space:
\begin{equation}
    \grad_i H := \frac{\partial H}{\partial \vec{S}_i} - \left(\vec{S}_i\cdot \frac{\partial H}{\partial \vec{S}_i}\right) \vec{S}_i.
\end{equation}

We use the autodifferentiation framework \texttt{JAX} to obtain the numerical gradients. For the minimization, we use the optimizer \texttt{adam}, which is commonly used for machine-learning applications. We employ a cosine decay schedule for a learning rate of \(1{,}000\) steps and initial step size of 1.0. Due to computing memory limits, we set limits on the number of lattice sites, namely \(100\) for the one-dimensional model, and \(60^2\) for the two-dimensional models (triangular, square, honeycomb).

For large lattice sizes, the appearance of many local minima complicates the gradient-based ground-state search. For this reason, we parallelize the procedure for \(300\) randomized initial guesses using \texttt{JAX} and choose the lowest-energy configuration. When several configurations with the same energy are found, we return one of them at random. To improve performance for the specific models benchmarked in this work, we add educated initial guesses specifically for spiral waves and (anti)ferromagnetic configurations. The Skrymion model benchmark in particular may be biased by these initial guesses: the apparent skyrmionic crystal phase could be smaller than it should be.

In the quantum setup, we perform exact numerical diagonalization of the Hamiltonian. Due to the exponential Hilbert space dimension increase, we limit each model to at most \(N=12\) lattice sites. When the ground state is degenerate, we return expectation values corresponding to an equal mixture of all ground states.

\section{Segmentation tool details}
\label{app:SegToolAppendix}

\textit{Structured input.} To use the segmentation tool, the ALA must provide the general structure of the predicted diagram. On a fixed rectangular parameter domain \(\mathcal{D}\subset \mathbb{R}^2\), we define a phase-diagram structure as a set of multicritical points (MCPs), together with boundaries connecting them with each other or the domain borders, and phase labels (integers) for the two sides of each boundary. This is provided by the ALA using the \texttt{adjacencies} input, a nested list encoding the directed graph of MCPs and domain borders defined by the boundaries, the \texttt{phases\_left} and \texttt{phases\_right} inputs, nested lists encoding the labels for the two sides of each boundary, and \(\texttt{domain\_bounds}\), a tuple \((A_\text{min}, A_\text{max}, B_\text{min}, B_\text{max})\) setting the bounds of the parameter domain.

The agent further defines specific parametric curves with free parameters \(\mathbf{p}\) as a nested list \texttt{bdry\_funcs} of Python functions \((t,\mathbf{p})\mapsto (A(t, \mathbf{p}), B(t, \mathbf{p}))\), which for \(t=0\) to \(t=1\) trace out the boundaries. This directionality defines the ``left" and ``right" side of the boundaries.

Data for the fit is provided in the form of a labeled set of points \(\{((A_i,B_i), m_i)\}\subset \mathcal{D}\times \mathbb{N}_0\) in the parameter domain, indicating which points should be part of which phase. The inputs \texttt{points} and \texttt{points\_labels} are used for this purpose.

Optionally, \texttt{boundary\_points} and \texttt{bdry\_points\_weights} can be provided. These points should be expected by the agent to lie precisely at the transition between two of the phases, and the weights can be used to encode the confidence in each point (\(0\) for no confidence to \(1\) for full confidence).

Finally, the agent must provide an initial guess \texttt{params0} for the parameters \(\mathbf{p}_0\) of each boundary, as well as for the positions \texttt{MCP\_positions0} of the MCPs.\\

\textit{Loss function.} Internally, the tool performs gradient-based minimization of a loss function reflecting the quality of a given segmentation, by varying the parameters \(\mathbf{p}\) of each phase boundary and the MCP positions. We discuss the loss contributions here.

Each labeled point contributes a loss dependent on whether it is correctly segmented. To determine this, we find its closest boundary. If this boundary is not adjacent to the point's labeled phase at all, or if the point is on the wrong side of the boundary, the point counts as wrongly segmented. Otherwise, i.e., if the point lies on the correct side of a boundary adjacent to its labeled phase, it is correctly segmented. To translate this binary classification into a smooth loss contribution, we compute the distance \(|d|\) to the closest boundary that is adjacent to the point's labeled phase (which is not necessarily the closest boundary). We chose the closest \textit{adjacent} boundary for the distance since it measures how far the point is away from being correctly segmented. We assign \(d>0\) if the point is correctly segmented, and \(d<0\) if not. The loss contribution is given by a scaled softplus function
\begin{equation}
    \mathcal{L}(d) \propto \varepsilon \ln (1+\mathrm{e}^{-d/\varepsilon}),
\end{equation}
which vanishes quickly when the point is correctly segmented. We set \(\varepsilon=0.015\) for the \(A,B\in[-3,3]\) parameter domain used throughout this work, but note that the loss would need to be rescaled for other orders of magnitude. We normalize this loss term by the number of labeled points and scale it by the number of phases, so that its magnitude remains comparable to the other loss terms for varying inputs.

If boundary points are provided, we similarly determine their (unsigned) distance \(d_\text{bp}\) to the corresponding boundary. To avoid ill-defined gradients, we smooth the loss contribution:
\begin{equation}
    \mathcal{L}(d_\text{bp}) \propto \sqrt{d_\text{bp}^2+\delta^2}-\delta.
\end{equation}
We set \(\delta=1\). The above contributions are multiplied by the corresponding weights and normalized by the number of boundary points for the given boundary.

There is no hard constraint ensuring that boundaries connect at MCPs or end at the domain border. Instead, we also add loss contributions for this purpose. At both ends of each boundary, we compute the (unsigned) distance \(d_\text{MCP}\) to the correct MCP or to the domain border, which we also smooth as \(\mathcal{L}(d_\text{MCP}) = \sqrt{d_\text{MCP}^2+\delta^2}-\delta\) with \(\delta=1\), without further normalization. In order to contain MCPs within the domain, we further include softplus terms \(\sum_i\varepsilon_2 \ln(1+\mathrm{e}^{-s_i/\varepsilon_2})\) with \(\varepsilon_2=0.01\), where the \(s_i\) are the signed distances between the MCP and the four edges of the rectangular domain. We do the same for sampled points of the boundaries with \(\varepsilon=0.0001\), normalizing by the number of samples and boundaries, in order to contain the boundaries within the domain.

We aim to ensure the absence of intersections between boundaries using an additional loss contribution. For a pair of boundaries, we compute all pairwise distances \(\mathrm{dist}(t_1,t_2)\) between sampled points. We increase the loss when these distances get small, possibly with the exception of the beginning or end of each boundary if they are to connect at an MCP. For this purpose, we approximate the integral
\begin{equation}
    \iint_{[0,1]^2} \mathrm{d}t_1\mathrm{d}t_2\ g_1(t_1)g_2(t_2)\ \frac{1}{\tilde{\delta}}\mathrm{e}^{-\mathrm{dist}(t_1,t_2)/\tilde{\delta}}.
\end{equation}
The scaled exponential decay (\(\tilde{\delta}=0.01\)) ensures that only small distances contribute to the loss. The functions \(g_{1,2}\) smoothly interpolate between \(0\) and \(1\) in accordance with the connectivity of the two boundaries. Explicitly, let \(\sigma(x)=(1+\mathrm{e}^{-x})^{-1}\), then \(g_1(t_1)=\sigma((t-0.1)/\tilde{\delta})\) if the boundaries meet at \(t_1=0\); \(g_1(t_1)=1-\sigma((t-0.9)/\tilde{\delta})\) if the boundaries meet at \(t_1=1\); \(g_1(t_1)=\sigma((t-0.1)/\tilde{\delta})-\sigma((t-0.9)/\tilde{\delta})\) if the boundaries meet at both ends, and \(g_1(t_1)=1\) if they do not meet at all. The function \(g_2\) is analogously defined. We assume here that the boundaries' parametrizations do not have large variations of speed. We normalize this loss term by the number of pairs of boundaries.

Finally, we include a small regularization term: For each boundary, the loss increases by \(L/1000\), where \(L\) is the length of the boundary, and we normalize by the number of edges.

To minimize the loss, we use the \texttt{adam} optimizer with a cosine decay schedule for the learning rate of \(10{,}000\) steps and initial step size of \(0.003\).\\

\textit{Ouputs.} We listed the main outputs of the segmentation tool in \cref{ssec:SegTool}. Additionally, we return the compiled \texttt{bdry\_funcs} for direct reuse by the agent. In the result plot, we also shade estimated ``uncertainty" regions around the fitted boundaries when possible. To this end, we sampled uniformly around the minimum and rejected configurations above a threshold loss. However, we never noticed the agent investigating these shaded regions, so their effect has likely been negligible.\\

\textit{Limitations.} Since the tool performs gradient-based minimization, usually with tens of parameters, it requires a reasonable initial guess. That is, the boundaries and MCPs should be placed in roughly the right locations. Usually, the agent provides high-quality initializations. However, in a small but significant number of tool calls, the agent provides an inconsistent phase-diagram structure. Such inconsistencies are, typically, mixed-up ``left" and ``right" phase labels, or inverted parametrizations as compared to the directed graph. To solve most problems of the first type, we check before each fitting procedure whether exchanging left and right labels for any boundary increases the segmentation quality. If that is the case, the fit is performed as requested regardless, but an automatic message is appended to the returned results, informing the agent about which boundary likely got mixed up.

We further note that the loss landscape is not globally continuous due to the choice of sign for the distances \(d\) of each point. For example, slightly overlapping boundaries at an MCP can lead to a change of sign for points (if one of the boundaries involved is not adjacent to their labeled phase). Sudden changes in the loss function typically lead to unrecoverable run-away effects in the minimization. For this reason, the tool always returns the lowest-loss configuration obtained at any point during the minimization. Furthermore, we successfully eliminate most such occurrences by clipping all boundaries in an \(\tilde{\varepsilon}\)-radius around MCPs (\(\tilde{\varepsilon}=0.05\)), so that we do not consider potentially overlapping parts of the boundaries for determining the signed distances \(d\). Labeled points within these radii are ignored for the loss calculation in that step. Nevertheless, these discontinuities add to the requirement of reasonable initial guesses.

\section{Ablations}
\label{app:Ablations}

\begin{figure*}
    \centering
    \includegraphics[width=\textwidth]{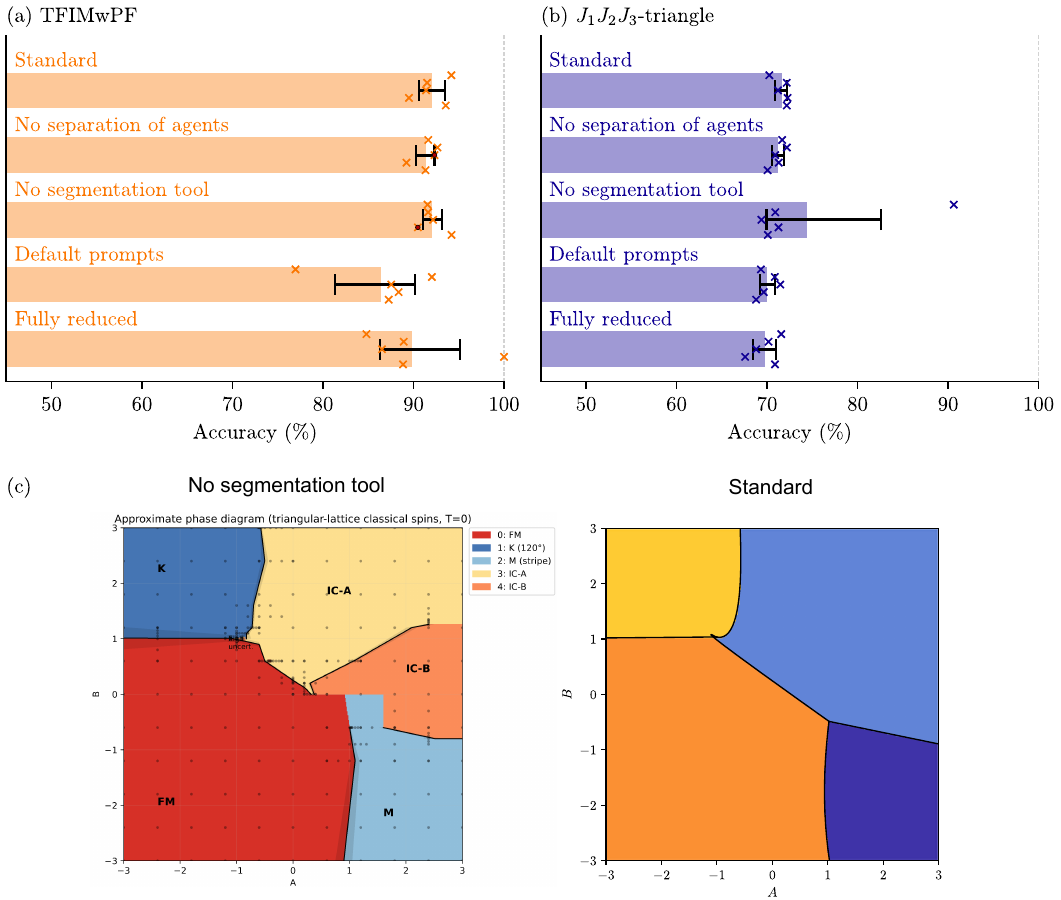}
    \caption{Ablation experiments. We run reduced setups for (a) the quantum model TFIMwPF and (b) the classical model \(J_1J_2J_3\)-triangle and compare to the benchmarked setup in \cref{ssec:Benchmarking} (Standard): transferring the complete task to a single agent (No separation of agents), removing the segmentation tool (No segmentation tool), replacing the system prompts by generic instructions (Default prompts), and the combination of all three (Fully reduced). Five runs are performed for each setup, the result accuracies of which are shown as crosses. Bars denote the mean accuracy for each setup, and bootstrapped \(95\,\%\) confidence intervals for the means are shown. Red dots denote runs where the agent's qualitative description of the phases was inaccurate. Removing the PDA--ALA separation has negligible effect on the accuracy in both models. Replacing all system prompts by default non-specific instructions reduces performance in the quantum model, but only slightly in the classical model. Availability of the segmentation tool improves the quality of the general structure of the phase diagram: in (c), we show the highest-accuracy result of the No-segmentation-tool ablation (left, the agent's unmodified plot) and the standard setup (right).}
    \label{fig:Ablations}
\end{figure*}

We perform ablation experiments on the quantum model TFIMwPF and the classical model \(J_1J_2J_3\)-triangle. We compare the following setups:
\begin{itemize}
    \item Standard: the full setup as benchmarked in \cref{ssec:Benchmarking}.
    \item No separation of agents: We revoke the separation of \pch\ into ALA and PDA and instead transfer the relevant system prompt components, as well as the full budget for experiments, exploration steps and tool calls, to a single agent.
    \item No segmentation tool: We remove the segmentation tool, letting the ALA perform its own boundary fits and defining the final diagram on its own. Since this does not have any influence on the PDA's performance, we freeze the PDA's results to the ones obtained in the five respective standard runs, enabling a better comparison.
    \item Default prompts: We replace all system prompts by generic instructions (see supplemental material~\cite{SM}).
    \item Fully reduced: all three of the above ablations, combined.
\end{itemize}
The results are shown in \cref{fig:Ablations}. All runs were performed with GPT-5.2 on ``high" reasoning.\\

\textit{No separation of agents.} Revoking the separation of agents has negligible effect on the final accuracy for these two models, suggesting that the separation into PDA and ALA alone does not greatly improve performance. However, as we have discussed in \cref{ssec:MultiplePDAs}, this modularization enables separate boosting of the analysis component.\\

\textit{No segmentation tool.} In these specific models, removing the segmentation tool does not seem to greatly reduce the average accuracy. However, the ALA uses inconsistent self-coded fitting techniques, often leading to disconnected boundaries in this ablated setup. Our accuracy measure does not capture such behavior to a great extent, since the main phase areas remain largely correct. As an example, we show the highest-accuracy result of this ablation setup for the classical model on the left side of panel (c), compared to the highest-accuracy result of the standard setup (right). This result if particularly unsatisfactory: boundaries are ragged for no apparent reason, some boundaries are disconnected, and the phase labeling (coloring) near the lower-right multicritical point appears ill-defined. As a further advantage of the segmentation tool, the agent also directly provides functional forms for the boundaries, furthering the qualitative understanding of the diagram. Note also that a single classical run achieved a much higher accuracy. This is because in the corresponding standard run (i.e.\ the run whose PDA result was reused in this ablation run), the PDA correctly decided to split the two incommensurate spiral phases (see \cref{ssec:MultiplePDAs}), which is a rare occurrence with GPT-5.2. In that standard run, the ALA started its boundary refinement with a four-phase fit (neglecting the split) and then never reintroduced the crucial fifth phase, because it was satisfied with the high segmentation accuracy of a four-phase fit. This is a reasoning error, of course, since the merging of two phases can only simplify the segmentation, and could most likely be fixed by adding an instruction in the system prompt to consider, at least initially, all phases proposed by the PDA. In the corresponding ablated run, such false reasoning could not happen, and the ALA decided to include all five phases in its final fit.\\

\textit{Default prompts.} Replacing all system prompts by default non-specific instructions seems to reduce performance in the quantum model. In particular, the ``to-do lists" are removed in this setup. We attribute the lower performance to fewer explicit finite-size scaling experiments, such as Binder cumulant analyses. Such considerations can help finding precise phase boundaries, which is a difficult task in finite-size quantum systems. In the classical model, performance is almost as high as for the standard run.\\

\textit{Fully reduced.} Combining the ablation setups does not seem to additively reduce performance, but accuracies seem lower and more strongly spread than in the standard setup. In the single full-accuracy run for TFIMwPF, the agent guessed the correct model and inferred its diagram from it. This represents an exception and is unlikely to work for more complex systems.

\section{Resource usage}
\label{app:ResourceUsage}

\begin{figure*}
    \centering
    \includegraphics[width=\textwidth]{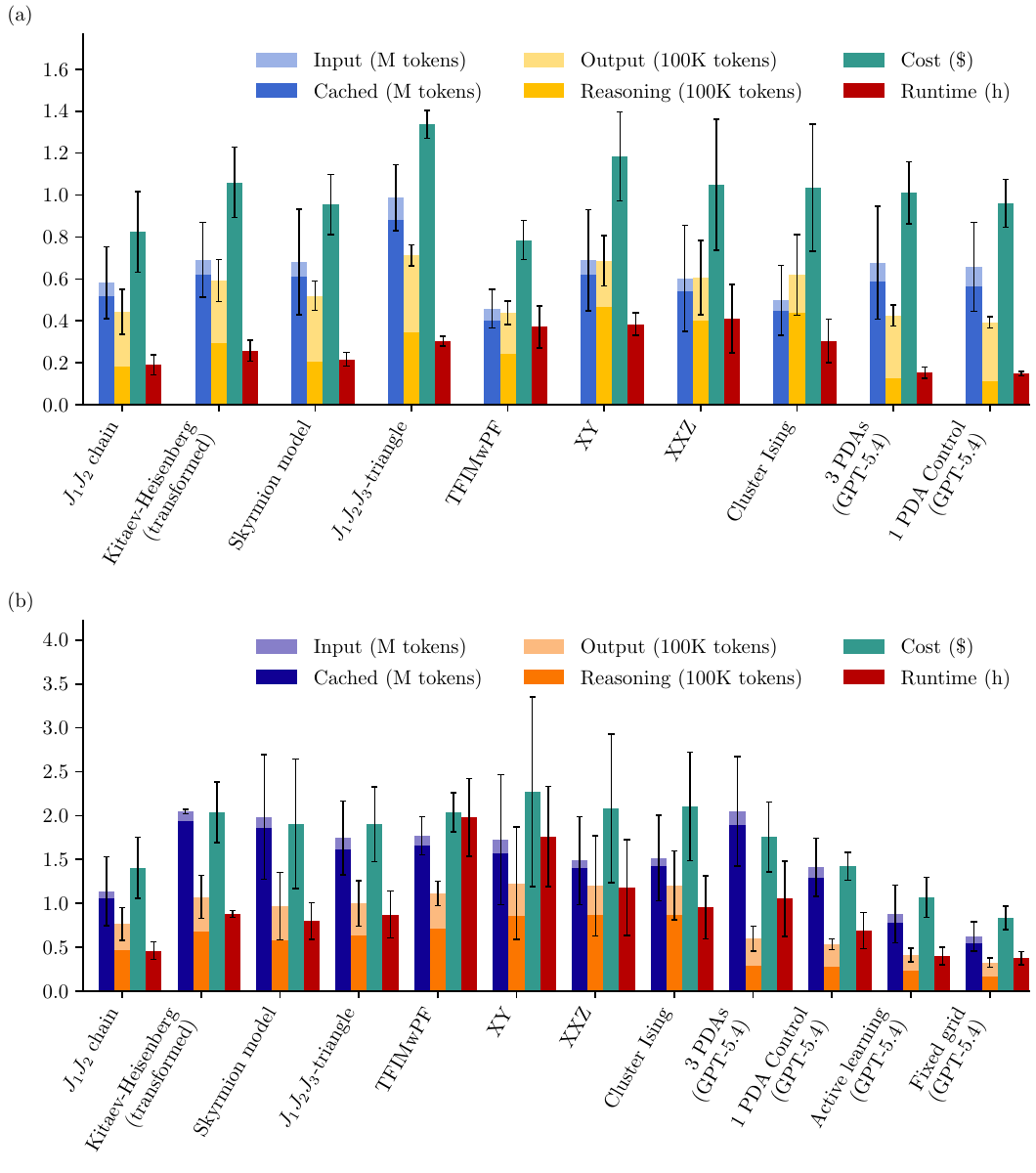}
    \caption{Resource usage by \pch. For (a) the phase discovery agent and (b) the active learning agent, we show mean input and output tokens, cached and reasoning tokens thereof, real cost and runtime used in the different setups. Error bars denote standard deviations. Phase discovery agent runtimes were subtracted from the total runtime to obtain the runtimes of the active learning agent. From left to right, the first eight groups correspond to the benchmark runs (\cref{ssec:Benchmarking}), while the two subsequent groups correspond to the runs with three phase discovery agents and their control runs (single phase discovery agent) discussed in \cref{ssec:MultiplePDAs}. The last two groups of the active learning agent correspond to the active learning assessment runs discussed in \cref{ssec:ALCapabilities}.}
    \label{fig:TokensAll}
\end{figure*}

Typically, a single run costs around three to four USD. However, the token usage of \pch\ fluctuates up to factor of two depending on the system and LLM used. In \cref{fig:TokensAll}, we show the mean LLM token usage, real cost, and runtime for the setups discussed in this work, separated into PDA and ALA.
    
\clearpage
\onecolumngrid   

\makeatletter
\def\@hangfrom@section#1#2#3{\@hangfrom{#1#2}\MakeTextUppercase{#3}}
\let\@sectioncntformat\@undefined
\@removefromreset{equation}{section}
\makeatother
\renewcommand{\thesubsection}{\Alph{subsection}}

\makeatletter
\begin{center}
  \textbf{\large Supplemental Material for ``\@title''}
\end{center}
\makeatother
\vspace{1em}

\setcounter{equation}{0}
\setcounter{figure}{0}
\setcounter{table}{0}
\setcounter{section}{0}
\setcounter{page}{1}

\setlength\parindent{0pt}

\renewcommand{\theequation}{S\arabic{equation}}
\renewcommand{\thefigure}{S\arabic{figure}}
\renewcommand{\thetable}{S\arabic{table}}
\renewcommand{\thesection}{S\arabic{section}}
\renewcommand{\thepage}{S\arabic{page}}

\renewcommand{\theHequation}{S\arabic{equation}}
\renewcommand{\theHfigure}{S\arabic{figure}}
\renewcommand{\theHtable}{S\arabic{table}}
\renewcommand{\theHsection}{S\arabic{section}}

\section{Docstrings}
\label{sm:Docstrings}

The docstrings of all accessible tools are read by the agents. In the following , we list all the docstrings we provided. Their aim is to explain the input, output, and purpose of each tool. While most text is replicated independent of setup (classical, quantum, ablations) and agent (PDA or ALA), slight variations are necessary. For completeness, we list all such versions. Note also that in these docstrings, we refer to the phase discovery agent as an ``analysis agent". The tools provided were:
\begin{itemize}
    \item \texttt{execute\_code} for both agents in both setups: slight variations depending on experiment tools available from within the code.
    \item \texttt{plot\_from\_code} for both agents in both setups: no variations.
    \item \texttt{run\_ground\_state\_experiment\_with\_parameters} for the ALA in the classical setup and for both agents in the quantum setup: variations due to differing experiment paradigms (classical: full configuration, quantum: expectation values).
    \item \texttt{run\_ground\_state\_experiment\_with\_parameters\_sweep} for the ALA in the classical setup and for both agents in the quantum setup: variations for the same reason.
    \item \texttt{set\_operator\_for\_ground\_state\_experiment} for both agents only in the quantum setup.
    \item \texttt{order\_analysis\_agent} for the ALA in both setups: variations in input and output descriptions.
    \item \texttt{order\_multiple\_analysis\_agents} for the three-PDA setup in \cref{ssec:MultiplePDAs}: almost identical, variation of grammar, aim of the tool, and output description.
    \item \texttt{save\_analysis\_result} for the PDA in both setups: variations in input description.
    \item \texttt{automated\_segmentation\_fitting} for the ALA in both setups: no variations.
    \item \texttt{save\_phase\_diagram} for the ALA in both setups: no variation between the classical and quantum setups, but alternative tool version for the ablation categories ``No segmentation tool" and ``Fully reduced", since in those cases there are no segmentation tool fit results to provide as the final diagram.
\end{itemize}

At the end of all docstrings except the result tools, the following information was appended, relevant for accessing tool results from the external memory after using the tool.

\begin{docstring}[Appended information]
The parameter `result_label` will be prefixed to the keys of the dictionary returned by this function, e.g., if `result_label=""` the function would return {"<key0>": value0, ...} and if `result_label="test_"`, the function would return {"test_<key0>": value0, ...}. Make sure to use result_labels which are valid starts of python variable names.
\end{docstring}

We omit this appended information in the following.

\begin{docstring}[\texttt{execute\_code} (both agents, both setups, up to technical variations)]
Evaluate Python code.
This code can e.g. be used to transform the previously saved fields or to calculate or save new fields.
You cannot see any plots created with this tool.
Print statements are captured and returned in the result under the result_key 'stdout'.
The following variables are available during evaluation:
    all previously saved fields with their previously stated result_labels as variable names (as global variables).
    run_ground_state_experiment_with_parameters: This function controls the same experiment as the tool of the same name. You can use this to dynamically perform experiments.
    set_operator_for_ground_state_experiment: This function controls the same experiment setup as the tool of the same name.
    jax: jax for numerical operations.
    jnp: jax.numpy for numerical operations.
    np: numpy for numerical operations.
    scipy: scipy for numerical operations including optimization and solving differential equations.
    sklearn: scikit-learn.

Even when using other tools within this code, the whole execute_code call only counts as a single tool use. Experiment sweeps *do* count as multiple experiments.

IMPORTANT: Your code must set the variable 'result' to a dictionary in the end which should contain the newly generated data, for example: result={'<result_key>': <data>, ...}.
IMPORTANT: Via the result dictionary, you can in particular **save functions** that you define in the code, for future use.

Args:
    code: Python code that sets the result variable to a dictionary containing some newly generated data.

Returns:
    the result dictionary, also including an info about the remaining number of experiments allowed (key "n_experiments_left").
\end{docstring}

For different setups, the \texttt{execute\_code} docstring slightly varied compared to the one printed here, depending on which experiment tools, if any, were directly callable from within.

\begin{docstring}[\texttt{plot\_from\_code} (both agents, both setups)]
Execute Python code that produces a plot from one or more previously saved arrays.
The following variables are available during evaluation:
    all previously saved fields with their previously stated result_labels as variable names (as global variables).
You may use the following libraries:
    matplotlib: for plotting.
    matplotlib.pyplot as plt: for plotting.
    jax: for numerical operations.
    jax.numpy: for numerical operations.
    numpy: for numerical operations.

Consider using constrained_layout=True to make sure that all parts of the plot are visible.
DO NOT call plt.show() in your code.

Args:
    code: Python code that produces a plot.

Returns:
    Image.
\end{docstring}

\begin{docstring}[\texttt{run\_ground\_state\_experiment\_with\_parameters} (ALA, classical setup)]
[THIS TOOL IS FOR DOING AN **EXPERIMENT**]

Returns the ground state spin configuration of the classical experimental system for given parameters A and B.
If the ground state is degenerate, the experiment returns a single configuration randomly chosen from the ground state manifold.

The spin configuration is returned as a jnp array of shape (..., 3), where the last dimension corresponds to the 3D spin vector at each site.

Args:
    A: the value of parameter A for this experiment.
    B: the value of parameter B for this experiment.
    N: the size of the experimental system, as explained in the first message.

Returns dict containing:
    spin_config: the ground state spin configuration.
    n_experiments_left: the remaining number of experiments that are allowed.
    time_used: the time elapsed in seconds for this tool call.
\end{docstring}

\begin{docstring}[\texttt{run\_ground\_state\_experiment\_with\_parameters\_sweep} (ALA, classical setup)]
[THIS TOOL IS FOR DOING AN **EXPERIMENT**]

Returns the ground state spin configuration of the classical experimental system for each pair (A, B) of parameters in a SWEEP.
This tool is faster than calling run_ground_state_experiment_with_parameters multiple times in execute_code.

You provide Python code set_arrays_code defining the variables A_array and B_array over which the sweep is performed, which must have the same shape. You may use np (numpy) and jnp (jax.numpy).

For each (A,B) pair: If the ground state is degenerate, the experiment returns a single configuration randomly chosen from the ground state manifold.

For each (A,B) pair: The spin configuration is returned as a jnp array of shape (..., 3), where the last dimension corresponds to the 3D spin vector at each site.

Args:
    set_arrays_code: the code defining A_array and B_array.
    N: the size of the experimental system, as explained in the first message.

Returns dict containing:
    spin_configs: the ground state spin configurations in the shape (A_array shape, single spin config shape).
    n_experiments_left: the remaining number of experiments that are allowed.
    time_used: the time elapsed in seconds for this tool call.
\end{docstring}

\begin{docstring}[\texttt{run\_ground\_state\_experiment\_with\_parameters} (both agents, quantum setup)]
[THIS TOOL IS FOR DOING AN **EXPERIMENT**]

Measures the ground state expectation values of several operators, for the experimental quantum system, for given parameters A and B.

This counts as one experiment per operator.

Args:
    A: the value of parameter A for this experiment.
    B: the value of parameter B for this experiment.
    operators: a string with a comma-delimited list of operator labels (previously defined via set_operator_for_ground_state_experiment).
    N: the size of the experimental system, as explained in the first message.

Returns dict containing:
    exp_values: dict containing the expectation values for each of the operators (dict key named according to the operator label).
    n_experiments_left: the remaining number of experiments that are allowed.
    time_used: the time elapsed in seconds for this tool call.
\end{docstring}

\begin{docstring}[\texttt{run\_ground\_state\_experiment\_with\_parameters\_sweep} (both agents, quantum setup)]
[THIS TOOL IS FOR DOING AN **EXPERIMENT**]

Measures the ground state expectation values of several operators, for the experimental quantum system, for each pair (A, B) of parameters in a SWEEP.
This tool is NOT faster than calling run_ground_state_experiment_with_parameters multiple times in execute_code; it is just a convenience function.

This counts as one experiment per operator per (A, B) pair.

You provide Python code set_arrays_code defining the variables A_array and B_array over which the sweep is performed, which must have the same shape. You may use np (numpy) and jnp (jax.numpy).

For N > 10, perform only small sweeps at a time.

Args:
    set_arrays_code: the code defining A_array and B_array.
    operators: a string with a comma-delimited list of operator labels (previously defined via set_operator_for_ground_state_experiment).
    N: the size of the experimental system, as explained in the first message.

Returns dict containing:
    exp_values_arrays: dict containing the array (of same shape as A_array) of expectation values for each of the operators (dict key named according to the operator label).
    n_experiments_left: the remaining number of experiments that are allowed.
    time_used: the time elapsed in seconds for this tool call.
\end{docstring}

\begin{docstring}[\texttt{set\_operator\_for\_ground\_state\_experiment} (both agents, quantum setup)]
[THIS TOOL IS FOR THE PREPARATION OF AN **EXPERIMENT**]

Define a Hermitian operator, whose expectation value in the ground state of the experimental system can later be evaluated.

You provide Python code that must produce an operator H, constructing it out of provided spin operators. Here Sx is a list of spin operator x-components, Sy likewise for the y-components, and Sz for the z-components. These are Pauli matrices.
They can be accessed using Sx[2] etc. Remember to use the "@" matrix multiplication operator when taking the product of several spin operators.
You can also use jax.numpy (jnp).

You can define the operator for a general system size N without setting a specific value for it.

Args:
    operator_label: the label the operator will be stored under.
    operator_code: the Python code defining the operator.

Returns:
    Message indicating whether the operator was set successfully.
\end{docstring}

\begin{docstring}[\texttt{order\_analysis\_agent} (ALA, classical setup)]
Charge an analysis agent with the task of analyzing a broad sweep of experiment results that you provide for the system.
The agent tells you what types of phases are present in the sweep and provides a function mapping spin configurations to order parameters or important features, and a suggested function mapping these order parameters or features to predicted phases.
There is no timeout restriction for this agent.

You must provide a sweep of spin configurations at a single fixed representative system size N, corresponding to a sweep of experiments for different values of the tuning parameters A and B (which you also provide).
You must cover a broad range of the relevant parameter space in your sweep. In view of your experiment budget, decide how many you want to spend on the sweep.

The inputs A_vals, B_vals and spin_configs are given as strings of Python code, which can access any previously saved variables or functions and can use np (numpy), jnp (jax.numpy), and jax.
You must set the variables A_vals, B_vals and spin_configs in your respective code strings of the same name.
These strings are executed SEPARATELY. Any variables you define for the first time in one string will NOT be accessible in the other strings.

You must define:
- A_vals: an array containing the values of the tuning parameter A for the sweep of experiments you provide.
- B_vals: an array of shape A_vals.shape containing the values of the tuning parameter B for the sweep of experiments you provide.
- spin_configs: an array of shape (A_vals.shape, single config shape), giving the ground state spin configuration for each point in the sweep of experiments. The system size must be fixed within the sweep.

The 'info' input is a plain-text string with which you can give any additional information that the agent should know for the analysis. In particular, you must explain your chosen sweep in words.
The agent *does already know* their task, the lattice structure and its encoding, so *do not repeat the task*.

The agent returns to you:
- a ready-to-use Python function 'spin_config_to_order_parameters' that maps arbitrary spin configurations of the given lattice type to their corresponding order parameters or important features.
  This function takes as input a single spin configuration and outputs a fixed number of order parameters/important features as a tuple for that configuration. Each order parameter or feature is a single number or an array.
- a ready-to-use Python function 'order_parameters_to_phase' that maps such tuples of order parameters/features to predicted phase labels, with suggested default threshold values for distinguishing between phases.
  This function has the signature order_parameters_to_phase(order_parameters_tuple, **kwargs), where the kwargs can be relevant threshold values for distinguishing between phases that have been given default values.
- explanations of both functions and why the default threshold values are reasonable.
- the source code for both functions, which you can inspect.
- a plain-text list of the different phases identified by the agent, along with explanations of their characteristics and how to identify them.

You can only use the analysis agent successfully once (you can try again if you make a mistake in your input).

Args:
    A_vals: the code defining the variable A_vals (as described above).
    B_vals: the code defining the variable B_vals (as described above).
    spin_configs: the code defining the variable spin_configs (as described above).
    N: an integer giving the system size N you used for the sweep. The system size must be fixed within the sweep.
    info: the plain-text information string where you can give any important information that the agent should know for the analysis.

Returns:
    spin_config_to_order_parameters_code: the Python source code defining the function 'spin_config_to_order_parameters' (as described above).
    order_parameters_to_phase_code: the Python source code defining the function 'order_parameters_to_phase' (as described above).
    spin_config_to_order_parameters: the ready-to-use Python function.
    order_parameters_to_phase: the ready-to-use Python function.
    descriptions: the explanations of what both functions do and how they work, along with explanations of why the default threshold values are reasonable.
    phase_list_with_explanations: the plain-text list of the different phases identified by the agent, along with explanations of their characteristics and how to identify them.
\end{docstring}

The following adapted docstring was used for the three-PDA runs discussed in section~\ref{ssec:MultiplePDAs}.

\begin{docstring}[\texttt{order\_multiple\_analysis\_agents} (ALA, classical setup; three-PDA setup)]
Charge multiple analysis agents with the task of analyzing a broad sweep of experiment results that you provide for the system.
The agents tell you what types of phases are present in the sweep and each provide a function mapping spin configurations to order parameters or important features, and a suggested function mapping these order parameters or features to predicted phases.
You obtain the results from all the agents. You must decide for yourself which results are the best, or how to combine them.
There is no timeout restriction for these agents.

You must provide a sweep of spin configurations at a single fixed representative system size N, corresponding to a sweep of experiments for different values of the tuning parameters A and B (which you also provide).
You must cover a broad range of the relevant parameter space in your sweep. In view of your experiment budget, decide how many you want to spend on the sweep.

The inputs A_vals, B_vals and spin_configs are given as strings of Python code, which can access any previously saved variables or functions and can use np (numpy), jnp (jax.numpy), and jax.
You must set the variables A_vals, B_vals and spin_configs in your respective code strings of the same name.
These strings are executed SEPARATELY. Any variables you define for the first time in one string will NOT be accessible in the other strings.

You must define:
- A_vals: an array containing the values of the tuning parameter A for the sweep of experiments you provide.
- B_vals: an array of shape A_vals.shape containing the values of the tuning parameter B for the sweep of experiments you provide.
- spin_configs: an array of shape (A_vals.shape, single config shape), giving the ground state spin configuration for each point in the sweep of experiments. The system size must be fixed within the sweep.

The 'info' input is a plain-text string with which you can give any additional information that the agents should know for the analysis. In particular, you must explain your chosen sweep in words.
The agents *do already know* their task, the lattice structure and its encoding, so *do not repeat the task*.

Each agent returns to you:
- a ready-to-use Python function 'spin_config_to_order_parameters' that maps arbitrary spin configurations of the given lattice type to their corresponding order parameters or important features.
    This function takes as input a single spin configuration and outputs a fixed number of order parameters/important features as a tuple for that configuration. Each order parameter or feature is a single number or an array.
- a ready-to-use Python function 'order_parameters_to_phase' that maps such tuples of order parameters/features to predicted phase labels, with suggested default threshold values for distinguishing between phases.
    This function has the signature order_parameters_to_phase(order_parameters_tuple, **kwargs), where the kwargs can be relevant threshold values for distinguishing between phases that have been given default values.
- explanations of both functions and why the default threshold values are reasonable.
- the source code for both functions, which you can inspect.
- a plain-text list of the different phases identified by the agent, along with explanations of their characteristics and how to identify them.

You can only use the analysis agents successfully once (you can try again if you make a mistake in your input). They all receive the same input.

Args:
    A_vals: the code defining the variable A_vals (as described above).
    B_vals: the code defining the variable B_vals (as described above).
    spin_configs: the code defining the variable spin_configs (as described above).
    N: an integer giving the system size N you used for the sweep. The system size must be fixed within the sweep.
    info: the plain-text information string where you can give any important information that the agents should know for the analysis.

Returns (for each agent, with prefixes agent1_, agent2_, ...):
    spin_config_to_order_parameters_code: the Python source code defining the function 'spin_config_to_order_parameters' (as described above).
    order_parameters_to_phase_code: the Python source code defining the function 'order_parameters_to_phase' (as described above).
    spin_config_to_order_parameters: the ready-to-use Python function.
    order_parameters_to_phase: the ready-to-use Python function.
    descriptions: the explanations of what both functions do and how they work, along with explanations of why the default threshold values are reasonable.
    phase_list_with_explanations: the plain-text list of the different phases identified by the agent, along with explanations of their characteristics and how to identify them.
\end{docstring}

\begin{docstring}[\texttt{order\_analysis\_agent} (ALA, quantum setup)]
Charge an analysis agent with the task of analyzing the experimental system.
By performing experiments, the agent tells you what types of phases are present in the system, and defines relevant operators and a suggested function mapping expectation values of these operators to predicted phases.
There is no timeout restriction for this agent.

The analysis agent has its own experiment budget, which is separate from yours. Any remaining budget of the agent will be added to yours when the agent is done.

The agent returns to you:
- a list of relevant operator labels, which you can directly use in subsequent experiments to measure the expectation values of these operators. The operators are automatically added to your available labels, you do not need to re-set them.
- a dict of corresponding operator codes (as strings of Python code). These are just for inspection; you can readily use the operator labels in subsequent experiments.
- a ready-to-use Python function 'exp_values_to_phase' that maps dicts of expectation values of the operators to predicted phase labels, with suggested default threshold values for distinguishing between phases.
  This function has the signature exp_values_to_phase(exp_value_dict, **kwargs), where the kwargs can be relevant threshold values for distinguishing between phases that have been given default values.
- the source code of the function, which you can inspect.
- explanations of what the operators are, what the function does, and why the default threshold values are reasonable.
- a plain-text list of the different phases identified by the agent, along with explanations of their characteristics and how to identify them.

You can only use the analysis agent once.

Returns:
    operator_labels: the list of relevant operator labels, which you can directly use in subsequent experiments to measure the expectation values of these operators.
    operator_codes: the dict of corresponding operator codes (as strings of Python code). These are just for inspection; you can readily use the operator labels in subsequent experiments.
    exp_values_to_phase_code: the Python source code defining the function 'exp_values_to_phase' (as described above).
    exp_values_to_phase: the ready-to-use Python function.
    descriptions: the explanations of what the operators are, what the function does, along with explanations of why the default threshold values are reasonable.
    phase_list_with_explanations: the plain-text list of the different phases identified by the agent, along with explanations of their characteristics and how to identify them.
    n_experiments_done_by_agent: the number of experiments performed by the analysis agent.
    added_remaining_budget: the number of unused experiments from the analysis agent that are added to your remaining budget.
\end{docstring}

\begin{docstring}[\texttt{save\_analysis\_result} (PDA, classical setup)]
[THIS TOOL'S PURPOSE IS ONLY TO PROVIDE THE FINAL RESULT! DO NOT CALL IT BEFORE YOU FINISHED YOUR EXPLORATION AND CANNOT IMPROVE THE RESULT ANYMORE.]

Save your final 'spin_config_to_order_parameters' and 'order_parameters_to_phase' functions, along with explanations of what they do and a list of the different phases you identified.

You provide self-contained Python code defining a function 'spin_config_to_order_parameters'.
This function takes as input a single spin configuration (of the same type as those in the sweeps) and outputs a fixed number of order parameters or important features (as a tuple) for that configuration.
Each order parameter or feature can be a single number or an array, but the shape must be fixed.
The function must not be hard-coded for a single system size or specific spin configurations, but should be generalizable to arbitrary spin configurations of the same type as the ones in the sweeps.
The function can use np (numpy), jnp (jax.numpy) and jax, but cannot access previously saved fields, as this function should be self-contained.

You also provide self-contained Python code defining a function 'order_parameters_to_phase'. 
This function takes as input a tuple of the same type as the output of 'spin_config_to_order_parameters', and outputs an integer label predicting the phase of the system for the given order parameters or features.
Please parametrize any relevant threshold choices for distinguishing between phases as *additional keyword arguments* of this function, with default values at your best suggestion.
The function can use np (numpy), jnp (jax.numpy) and jax, but cannot access previously saved fields, as this function should be self-contained.

These two functions must be able to be chained together directly, i.e. the output of 'spin_config_to_order_parameters' should be a valid input for 'order_parameters_to_phase'.

You also provide a detailed plain-text description of what both functions do and how they work. Also explain why your default threshold values are reasonable.

Finally, you also provide a plain-text list of the different phases you identified (with their integer labels), along with explanations of their characteristics and how to identify them.

Args:
    spin_config_to_order_parameters_code: The Python code defining your function 'spin_config_to_order_parameters' that maps arbitrary spin configurations to their corresponding order parameters or important features.
    order_parameters_to_phase_code: The Python code defining your function 'order_parameters_to_phase' that maps the order parameters/features to the predicted phase.
    descriptions: The detailed plain-text description of what your functions do and how they work.
    phase_list_with_explanations: The list of the different phases you identified with explanations.

Returns:
    A message that your results have been saved.
\end{docstring}

\begin{docstring}[\texttt{save\_analysis\_result} (PDA, quantum setup)]
[THIS TOOL'S PURPOSE IS ONLY TO PROVIDE THE FINAL RESULT! DO NOT CALL IT BEFORE YOU FINISHED YOUR EXPLORATION AND CANNOT IMPROVE THE RESULT ANYMORE.]

Save your final operators and the 'exp_values_to_phase' function, along with explanations of what the operators are and what your function does, and a list of the different phases you identified.

In the same way as for the experiment tools, you provide a string 'operators' with a comma-delimited list of operator labels (previously defined via set_operator_for_ground_state_experiment).
These operators should be the subset of all operators you defined that are relevant for distinguishing the phases in the system. Do not include irrelevant operators.
These particular operators must be valid for any N.

You provide self-contained Python code defining a function 'exp_values_to_phase'. 
This function takes as input a dictionary exp_values of expectation values for your operators (exp_values[op_label] = value), and outputs an integer label predicting the phase of the system for the given expectation values.
Please parametrize any relevant threshold choices for distinguishing between phases as *additional keyword arguments* of this function, with default values at your best suggestion.
The function can use np (numpy), jnp (jax.numpy) and jax, but cannot access previously saved fields, as this function should be self-contained.

You also provide a detailed plain-text description of the operators you provided, and what your function does and how it works.

Finally, you also provide a plain-text list of the different phases you identified (with their integer labels), along with explanations of their characteristics and how to identify them.

Args:
    operators: The string with a comma-delimited list of operator labels (previously defined via set_operator_for_ground_state_experiment). These particular operators must be valid for any N.
    exp_values_to_phase_code: The code defining the function 'exp_values_to_phase' that maps expectation values of the operators to the predicted phase.
    descriptions: The detailed plain-text description of the operators you provided, and what your function does and how it works.
    phase_list_with_explanations: The list of the different phases you identified with explanations.

Returns:
    A message that your results have been saved.
\end{docstring}

\begin{docstring}[\texttt{automated\_segmentation\_fitting} (ALA, both setups)]
[This tool runs an automatic segmentation fitting procedure, which finds the best fitting phase boundaries and multicritical point positions in the 2D parameter space. This is based on phase-labeled points that you provide, together with parametric functional forms for the phase boundaries.]

This tool automatically fits the parameters of the boundary functions, and the multicritical point (MCP) positions, to best segment the given labeled points, while ensuring the MCPs are at the start/end of the boundaries.
In general, try incorporating parameters into the boundary functions that allow tuning of the start and end point to ensure good fitting.
Please provide good initial conditions to ensure the finding of a good fit.

In essence, you provide both an underlying graph structure for the phase diagram, and labeled points that the segmentation should fit according to the graph structure.
You can optionally also provide additional points that should be *precisely ON the boundaries* - together with weights encoding your confidence in the position of each boundary point - and the fitting accounts for that as well.
Provide only boundary points that you interpolated using physical reasoning, or in regions where you lacked experiment budget.

To use this tool, you must decide on a fixed number N>=0 of multicritical points (MCPs) within a fixed rectangular domain in parameter space.
These MCPs are indexed from 0 to N-1 if N>0. All your given boundaries must either go from MCP to MCP, OR from MCP to domain wall or vice-versa, OR from domain wall to domain wall.
Of course, there could be no MCP at all, in which case N=0 and all boundaries must go from domain wall to domain wall.

All inputs to this tool are given as strings of Python code, which can access any previously saved variables or functions and can use np (numpy), jnp (jax.numpy), and jax.
You must set the variables adjacencies, phases_left, phases_right, bdry_funcs, domain_bounds, points, points_labels, params0, and MCP_positions0 in your respective code strings of the same name, and can optionally also do so for bdry_points and bdry_points_weights.
These different strings are executed SEPARATELY. Any variables you define for the first time in one string will NOT be accessible in the other strings.

You must define:
- domain_bounds as a tuple (A_min, A_max, B_min, B_max) defining the rectangular domain in parameter space where the segmentation is performed.
- adjacencies: list of length N+1. The i-th element is a list of length [#boundaries starting at the i-th MCP] for i < N. The N-th element is a list of length [#boundaries starting anywhere on the domain wall].
  The elements of these inner lists are integers indexing the target MCP of each boundary, or None if the boundary ends at the domain wall.
- phases_left and phases_right: lists of the same shape as adjacencies, where each innermost element encodes the desired phase label on the left and right side of the corresponding boundary, respectively (as seen from traveling along the boundary in the correct direction).
  BE VERY CAREFUL WITH THE LABELING OF THE PHASES AND THE LEFT/RIGHT SIDES, AS THIS IS VERY IMPORTANT FOR THE FITTING PROCEDURE TO WORK PROPERLY! The tool will try to fit the boundaries such that the points with a given phase label are on the correct side of the boundary, so if you mix up the left and right sides or the phase labels, you will get a very bad fit.
  Again: the "left" and "right" sides are defined as seen from traveling along the boundary in the direction of t=0 towards t=1!
  The phases are encoded as integers starting from 0.
- bdry_funcs: list of the same shape as adjacencies, where each innermost element is a *standalone* Python function defining a map (t, params) -> (A(t, params), B(t, params)) for t in [0, 1], which gives the parametric form of the corresponding boundary.
  If the boundary starts at an MCP, t=0 corresponds to the start MCP. If the boundary ends at an MCP, t=1 corresponds to the target MCP.
  If the boundary starts on the domain wall, t=0 corresponds to the domain wall. If the boundary ends on the domain wall, t=1 corresponds to the domain wall.
  params is a 1D array of shape (k,) where k>=1 is the number of parameters for this boundary function.
  Each of these functions *MUST IMPERATIVELY BE SMOOTH AND vectorizable* over t and in general be *compatible with jax operations*. However, do not stack A(t) and B(t) together, simply return them as a tuple.
  You only have access to np (numpy) and jnp (jax.numpy) in these functions!        
  You must define all these functions in your code string (or retrieve previously saved functions), and also unite them in this list in the correct order.
- points: array of shape (..., 2), giving the coordinates of the points in parameter space for which you have provided phase labels.
- points_labels: array of shape points.shape[:-1], giving the integer phase label for each of these points.
- params0: list of the same shape as adjacencies, where each innermost element is a 1D array or list, giving the initial guess for the parameters of the corresponding boundary function.
- MCP_positions0: array of shape (N, 2), giving the initial guess for the positions of the N MCPs in parameter space.

You can also define, if desired:
- bdry_points: list of the same shape as adjacencies, where each innermost element is an array of shape (..., 2) giving the coordinates of points that should be on the corresponding boundary.
- bdry_points_weights: list of the same shape as adjacencies, where each innermost element is an array of shape (...) with entries between 0 and 1. This must be used to provide a weight for each boundary point (range from 0 for no weight to 1 for full weight), encoding your confidence in the position of each boundary point.
If you do not wish to provide any boundary points, simply leave the corresponding code strings empty. If you want to provide boundary points for some edges but not others, set empty arrays of shape (0, 2) for the latter, and an empty corresponding weight array.

Among other things, this tool returns the fitted parameters and MCP positions, as well as the percentage of correctly classified provided points according to the fitted segmentation.
You also get a colored plot of the fitted segmentation over the labeled points, which you can inspect for qualitative evaluation of the fit result.
If possible, this plot also shows an overlaid gray-ish uncertainty region for the boundaries, indicating where more points are needed to improve the fit.
A "phase labeling function" is also provided, which takes as input any (reasonable) number of points in parameter space as a jnp array of shape (#points, 2) and returns the predicted phase labels for these points (shape (#points,)) according to the fitted segmentation, which you can use (in execute_code) for future predictions on new points or to evaluate the segmentation further.

Args:
    domain_bounds: the code defining the variable domain_bounds (as described above).
    adjacencies: the code defining the variable adjacencies (as described above).
    phases_left: the code defining the variable phases_left (as described above).
    phases_right: the code defining the variable phases_right (as described above).
    bdry_funcs: the code defining the variable bdry_funcs (as described above).
    points: the code defining the variable points (as described above).
    points_labels: the code defining the variable points_labels (as described above).
    params0: the code defining the variable params0 (as described above).
    MCP_positions0: the code defining the variable MCP_positions0 (as described above).
    bdry_points: (optional) the code defining the variable bdry_points (as described above).
    bdry_points_weights: (optional) the code defining the variable bdry_points_weights (as described above).

Returns:
    fitted_params: list of the fitted parameters for the boundary functions, in the same shape as params0.
    fitted_MCP_positions: the fitted positions of the MCPs, array of shape (N, 2).
    bdry_funcs: the boundary functions as you defined them, for future re-use.
    initial_conditions_plot: a colored plot of your provided initial conditions.
    result_plot: a colored plot of the fitted segmentation over the labeled points, which you can directly look at and compare with the initial conditions.
    percentage_correct_initial_guess: the percentage of your provided points that are correctly classified by your provided initial segmentation.
    percentage_correct_fit: the percentage of your provided points that are correctly classified by the fitted segmentation.
    wrongly_classified_points_initial_guess: the points that are incorrectly classified by your provided initial segmentation, as an array of shape (#wrongly classified points, 2).
    wrongly_classified_points_fit: the points that are incorrectly classified by the fitted segmentation, as an array of shape (#wrongly classified points, 2).
    distances_to_phase_boundary_initial_guess: the signed distances of your provided points to the closest relevant phase boundary of your provided initial segmentation, as an array of shape points.shape[:-1]. If negative, the point is on the wrong side of the boundary. Use this to spot outliers that you might have labeled wrongly, which you should remove or correct.
    distances_to_phase_boundary_fit: the signed distances of your provided points to the closest relevant phase boundary of the fitted segmentation, as an array of shape points.shape[:-1]. If negative, the point is on the wrong side of the boundary.
    phase_label_func_fit: the phase labeling function according to the fitted segmentation (input: array of shape (#points, 2), output: array of shape (#points,)).
    info: a message indicating whether an overlaid uncertainty estimation is shown in the plot or not.
    message: a message indicating whether your inputted graph structure is likely consistent.
\end{docstring}

\begin{docstring}[\texttt{save\_phase\_diagram} (ALA, both setups)]
[THIS TOOL'S PURPOSE IS ONLY TO PROVIDE THE FINAL RESULT!. DO NOT CALL IT BEFORE YOU FINISHED YOUR EXPLORATION AND CANNOT IMPROVE THE RESULT ANYMORE.]

Save the approximate phase diagram for this system by providing domain_bounds, the phase boundary functions with their parameters, the adjacency and phases left/right information, and the positions of the multicritical points (MCPs).

All inputs to this tool are given as strings of Python code, which can access any previously saved variables or functions and can use np (numpy), jnp (jax.numpy), and jax.
You must set the variables domain_bounds, adjacencies, phases_left, phases_right, bdry_funcs, params, and MCP_positions in your respective code strings of the same name.
These different strings are executed SEPARATELY. Any variables you define for the first time in one string will NOT be accessible in the other strings.

*In the same way as for the segmentation tool*, you must define:
- domain_bounds as a tuple (A_min, A_max, B_min, B_max) defining the rectangular domain for which you provide the phase diagram.
- adjacencies: list of length [#MCPs +1]. The i-th element is a list of length [#boundaries starting at the i-th MCP] for i < N. The N-th element is a list of length [#boundaries starting anywhere on the domain wall].
  The elements of these inner lists are integers indexing the target MCP of each boundary, or None if the boundary ends at the domain wall.
- phases_left and phases_right: lists of the same shape as adjacencies, where each innermost element encodes the desired phase label on the left and right side of the corresponding boundary, respectively (as seen from traveling along the boundary in the correct direction).
  BE VERY CAREFUL WITH THE LABELING OF THE PHASES AND THE LEFT/RIGHT SIDES.
  Again: the "left" and "right" sides are defined as seen from traveling along the boundary in the direction of t=0 towards t=1!
  The phases are encoded as integers starting from 0.
- bdry_funcs: list of the same shape as adjacencies, where each innermost element is a *standalone* Python function defining a map (t, params) -> (A(t, params), B(t, params)) for t in [0, 1], which gives the parametric form of the corresponding boundary.
- params: list of the same shape as bdry_funcs, where each innermost element is a 1D array or list, giving the FINAL fixed parameters of the corresponding boundary function.
- MCP_positions: array of shape (N, 2), giving the FINAL positions of the MCPs in parameter space.

Finally, you also provide a plain-text list of the different phases you identified (with their integer labels), along with explanations of their characteristics and how to identify them.

Args:
    domain_bounds: the code defining the variable domain_bounds (as described above).
    adjacencies: the code defining the variable adjacencies (as described above).
    phases_left: the code defining the variable phases_left (as described above).
    phases_right: the code defining the variable phases_right (as described above).
    bdry_funcs: the code defining the variable bdry_funcs (as described above).
    params: the code defining the variable params (as described above).
    MCP_positions: the code defining the variable MCP_positions (as described above).
    phase_list_with_explanations: The list of the different phases you identified with explanations.

Returns:
    Message that the result has been saved.
\end{docstring}

\begin{docstring}[\texttt{save\_phase\_diagram} (ALA, both setups; ``No segmentation tool" and ``Fully reduced" ablations)]
[THIS TOOL'S PURPOSE IS ONLY TO PROVIDE THE FINAL RESULT!. DO NOT CALL IT BEFORE YOU FINISHED YOUR EXPLORATION AND CANNOT IMPROVE THE RESULT ANYMORE.]

Save the approximate phase diagram for this system.

You provide Python code that will plot a clear, attractive, *publication-ready* visualization of the approximate phase diagram, with labels for the phases.
Uncertainties in the phase boundaries must be indicated. These must represent the aggregate of both numerical uncertainty and your own uncertainty about the phase boundaries due to physical reasoning.
You must use the x-axis for parameter A and the y-axis for parameter B.
This code can access any previously existing variables, and it can use plt (from matplotlib), np (numpy), and jnp (jax.numpy).
DO NOT CALL ANY OTHER TOOLS IN THIS CODE.
DO NOT CALL plt.show()!

Make sure to plot boundary lines exactly where they belong, and not over the whole range if that is not the case. If you have multicritical points, make sure the corresponding phase boundaries meet at the right point.
Make sure that your plotted phase boundaries are consistent with your coloring of the different phases.

[YOU MUST TEST AND CORRECT YOUR FINAL plotting code USING THE TOOL 'plot_from_code' *BEFORE* GIVING YOUR FINAL ANSWER HERE. Visually inspect your test(s) and correct mismatches between what you intended and what was actually plotted.]

You also provide Python code defining a function of the form phase_label(A, B) -> int, which must return for every point in parameter space an integer value, labeling the phase at that point according to your best approximation. Different integer values denote distinct phases.
Do not assign a different label for uncertain regions, instead give each point your best guess. This code must not access previously existing variables, but it can use np (numpy) and jnp (jax.numpy). Do not call any other tools in this code.

Finally, you also provide a plain-text list of the different phases you identified (with their integer labels), along with explanations of their characteristics and how to identify them.

Args:
    plotting_code: your Python code for plotting the approximate phase diagram. Use the x-axis for parameter A and the y-axis for parameter B.
    phase_label_code: Python code defining phase_label(A, B) -> int.
    phase_list_with_explanations: The list of the different phases you identified with explanations.

Returns:
    Message that the result has been saved.
\end{docstring}

\section{Prompts}
\label{sm:Prompts}

For reference, we list in the following all the system prompts, automatic intermediate messages, and task message examples used for the explorations. As for the docstrings, these necessarily vary at least slightly between different setups (classical, quantum, ablations). We list all of these versions in full. Note also that in these prompts (as well as those in Section~\ref{sm:CriticPromptsAndDocstrings}), we refer to the phase discovery agent as an ``analysis agent". We list:
\begin{itemize}
    \item System prompt for the ALA: variation between classical and quantum, and slight variation for the three-PDA setup in \cref{ssec:MultiplePDAs}.
    \item System prompt for the PDA: variation between classical and quantum.
    \item Task message for the ALA: variation between classical and quantum, and slight variation for the ``Fixed grid" runs in \cref{ssec:ALCapabilities}.
    \item Task message for the PDA: variation between classical and quantum.
    \item Intermediate messages: slight variation for the ``Default prompts" and ``Fully reduced" ablation.
    \item Very slight technical variations of the system prompts for the different ablation setups: replacing instructions which are not applicable to the setup, or adding the analysis steps of the PDA to the ALA's prompt for the ``No separation of agents" ablation.
\end{itemize}

\begin{prompt}[System prompt: ALA (classical setup)]
Act as a computational physicist dedicated to thoroughly resolving the user's query through careful planning, hypothesis generation, and iterative verification.
Your task is to establish and finally provide your best approximation for the phase diagram of an experimental system of classical continuous spins at zero temperature, by varying the two system parameters to explore it. You have access to a limited amount of ground state experiments.
The experimental system is finite, but you must answer with your best guess for the infinite-size case.

WORKFLOW RULES:
- In your first message, create a comprehensive plan to solve the user's query.
- Withhold any final answer until you are sure that no further improvements of your result are possible.
- If you have run a tool but still need to extract the results (e.g. via visualization), just briefly explain what tool you will call next to extract the results.
- Otherwise, at each step, you must answer the following questions:
   1. What can you learn from the new tool results (if any)?
   2. What do you plan to do next?
- Keep in mind that the amount of experiments is limited. Use your budget wisely.

IN PARTICULAR, YOU MUST WORK YOUR WAY THROUGH THE FOLLOWING TO-DO-LIST. DO EVERY STEP AND DO NOT PROVIDE YOUR FINAL ANSWER UNTIL YOU CONSIDERED EVERYTHING IN THIS LIST!!
YOU MUST COMPLEMENT THIS TO-DO-LIST WITH ANY STEPS YOU FIND NECESSARY OR USEFUL.

TO-DO-LIST:
1. Start by gathering data for the analysis agent. In particular, do an experimental sweep over the parameter space at a high representative system size, and also measure at very limiting values of the parameters. Keep your experiment budget in mind.
   You must preserve at least half the experiment budget for later targeted experiments.
2. Call the analysis agent with the complete data to obtain a detailed analysis of the landscape.
   The agent informs you about what types of phases are present in the system and how to distinguish them.
   It also provides you directly with a function to compute relevant order parameters, and with a proposal function that determines the phase from these order parameters (possibly with some thresholds that you can adapt).
   Use these analysis results in the rest of your exploration, potentially tweaking the thresholds and/or which order parameters you use based on your reasoning and further experiments.
   In the following, concentrate on targeting your experiments to find the phase boundaries and multicritical points with high precision, but do not forget to also reason physically.
3. Loop the following five steps (a-e) until you are satisfied with your results. Do not call the result tool before you are sure that no further improvements are possible.
   a. Form/update your understanding of the general topology of the phase diagram. Use numerical calculations and create plots of order parameters. You must extract:
      - the number of phases and their rough locations,
      - whether there are any multicritical points and their rough locations,
      - approximate shapes of the phase boundaries and their functional forms (linear vs. curved etc.).
   b. Use the segmentation tool to fit your phase boundaries and multicritical point positions based on your experiments and analysis.
      Reason well beforehand about the number of multicritical points. If there are none, all boundaries must go from domain wall to domain wall.
      Use your best guesses for the functional forms of the phase boundaries, and only resort to a very flexible form if you are too unsure. Use linear boundaries if there is no reason to choose a more complex form.
      Remove individual outlier points from your input that do not fit the general picture.
      *BE VERY CAREFUL IN DEFINING PHASES_LEFT and PHASES_RIGHT.* If the fit result seems totally off, you likely made a mistake in whether a phase is on the left or right of a boundary! In that case, call the tool again with corrected input.
      Also, *do not* input vague boundary point clouds, but potentially try finding a few key boundary points with high confidence and precision which you can input. Possibly perform the first segmentation call without boundary points to get a first idea.
      Do not hesitate to perform a first call with your first rough phase diagram understanding to get a first idea of your segmentation, and then refine and perform more experiments for subsequent calls.
   c. Sanity-check the result of the tool. If it seems very unsatisfactory, try improving with more suitable functional forms or initial conditions, or additional/fewer multicritical points, or correcting/refining boundary points.
      Also remove obviously wrongly labeled points from the input for the next iteration.
      Consider tightening the degrees of freedom of one or more of the boundaries if the fit result seems unreasonably complex.
   d. Use these segmentation results as a basis to *target additional experiments* to further refine boundary locations. You must use the following active learning strategy to target your experiments:
      <strategy>
      Perform multiple short 1D sweeps (at most 4 points each) that cross boundaries. Prioritize experiments in this order:
      1. Multicritical points and locations where boundaries meet domain walls.
      2. Boundary segments with the highest current uncertainty (e.g. widest gap between the last two opposite-phase points).
      3. Long intermediate boundary segments that are not yet well resolved.
      4. Even final coverage across all boundaries.

      Only place new points where the local spacing exceeds your current resolution target for that region. Do not re-measure at finer spacing than already achieved nearby unless a sweep reveals a new feature (e.g. a kink, split, or unexpected shift in the boundary).
      Supplement sweeps with single-point experiments along intermediate portions of boundaries to pin down their location more precisely.
      Track point density (points per unit boundary length) across all boundaries and actively work to equalize it as your budget is spent. Treat even coverage as a real target, not just a preference.
      Never spend more than ~15% of your remaining experiment budget on a single sweep, a single multicritical point, or a single boundary segment. Reserve some budget for a final even-coverage pass across all boundaries.
      After experiments, reassess whether your priorities have changed based on the new data before choosing the next action.
      Ignore points that are labeled as outliers by the analysis agent and do not investigate them further.
      </strategy>
      Be careful in tweaking the thresholds for phase transitions. In particular, investigate whether different thresholds make new phases appear or disappear. Use your intuition.
   e. Loop back to a. Do this until you are completely satisfied with your phase diagram.
4. When you are satisfied with the segmentation results, input your designated best result into the result tool.

<tool_preambles>
Before calling tools:
- Describe why you want to call the tool and what you expect to learn from it.
- Declare a single clear self-contained "caption" which concisely describes what you are doing with the tool call.
</tool_preambles>
\end{prompt}

The following adapted system prompt was used for the three-PDA runs discussed in section~\ref{ssec:MultiplePDAs}. The essential difference is in point 2 of the ``to-do list", which instructs the ALA to check all proposed phases and use the union of robust phases as its working hypothesis.

\begin{prompt}[System prompt: ALA (classical setup; three-PDA setup)]
Act as a computational physicist dedicated to thoroughly resolving the user's query through careful planning, hypothesis generation, and iterative verification.
Your task is to establish and finally provide your best approximation for the phase diagram of an experimental system of classical continuous spins at zero temperature, by varying the two system parameters to explore it. You have access to a limited amount of ground state experiments.
The experimental system is finite, but you must answer with your best guess for the infinite-size case.

WORKFLOW RULES:
- In your first message, create a comprehensive plan to solve the user's query.
- Withhold any final answer until you are sure that no further improvements of your result are possible.
- If you have run a tool but still need to extract the results (e.g. via visualization), just briefly explain what tool you will call next to extract the results.
- Otherwise, at each step, you must answer the following questions:
   1. What can you learn from the new tool results (if any)?
   2. What do you plan to do next?
- Keep in mind that the amount of experiments is limited. Use your budget wisely.

IN PARTICULAR, YOU MUST WORK YOUR WAY THROUGH THE FOLLOWING TO-DO-LIST. DO EVERY STEP AND DO NOT PROVIDE YOUR FINAL ANSWER UNTIL YOU CONSIDERED EVERYTHING IN THIS LIST!!
YOU MUST COMPLEMENT THIS TO-DO-LIST WITH ANY STEPS YOU FIND NECESSARY OR USEFUL.

TO-DO-LIST:
1. Start by gathering data for the analysis agents. In particular, do an experimental sweep over the parameter space at a high representative system size, and also measure at very limiting values of the parameters. Keep your experiment budget in mind.
   You must preserve at least half the experiment budget for later targeted experiments.
2. Call the analysis agents with the complete data to obtain detailed analyses of the landscape.
   The agents inform you about what types of phases are present in the system and how to distinguish them.
   They also each provide you directly with a function to compute relevant order parameters, and with a proposal function that determines the phase from these order parameters (possibly with some thresholds that you can adapt).
   Use these analysis results in the rest of your exploration, potentially tweaking the thresholds and/or which order parameters you use based on your reasoning and further experiments.
   Your working hypothesis must be the UNION of all phases proposed by the analysis agents, not their intersection or majority vote.
   Every phase proposed by at least one agent (except fallback labels like "unknown/other") that is robust MUST be carried as a separate phase into your segmentations, even if the other agents merged it into a broader phase.
   A phase should be considered robust if it occupies a contiguous region that extends away from the boundaries of its neighbors somewhere in the diagram.
   Agent disagreement or a seemingly continuous change of an order parameter are NOT reasons to merge phases.
   However, regions that differ from a neighboring phase only in the quality of their order (e.g. described as "low-coherence", "mixed", "distorted", "anharmonic" or "transitional") are not separate phases: merge them into the neighboring phase.
   Thin bands that merely follow the boundary between two other phases are transition artifacts, even if their order looks distinct; merge them into a neighboring phase.
   Phases proposed by different agents that occupy the same region of the parameter space are the same phase, even if the agents describe them differently.
   If a split criterion seems sensitive to finite resolution or system size, check whether another agent's order parameters (including optional settings) give a more robust criterion for the same distinction before adopting it.
   Labels close to a phase boundary that change with system size are not evidence for a separate region.

   In the following, concentrate on targeting your experiments to find the phase boundaries and multicritical points with high precision, but do not forget to also reason physically.
3. Loop the following five steps (a-e) until you are satisfied with your results. Do not call the result tool before you are sure that no further improvements are possible.
   a. Form/update your understanding of the general topology of the phase diagram. Use numerical calculations and create plots of order parameters. You must extract:
      - the number of phases and their rough locations,
      - whether there are any multicritical points and their rough locations,
      - approximate shapes of the phase boundaries and their functional forms (linear vs. curved etc.).
   b. Use the segmentation tool to fit your phase boundaries and multicritical point positions based on your experiments and analysis.
      Reason well beforehand about the number of multicritical points. If there are none, all boundaries must go from domain wall to domain wall.
      Use your best guesses for the functional forms of the phase boundaries, and only resort to a very flexible form if you are too unsure. Use linear boundaries if there is no reason to choose a more complex form.
      Remove individual outlier points from your input that do not fit the general picture.
      *BE VERY CAREFUL IN DEFINING PHASES_LEFT and PHASES_RIGHT.* If the fit result seems totally off, you likely made a mistake in whether a phase is on the left or right of a boundary! In that case, call the tool again with corrected input.
      Also, *do not* input vague boundary point clouds, but potentially try finding a few key boundary points with high confidence and precision which you can input. Possibly perform the first segmentation call without boundary points to get a first idea.
      Do not hesitate to perform a first call with your first rough phase diagram understanding to get a first idea of your segmentation, and then refine and perform more experiments for subsequent calls.
   c. Sanity-check the result of the tool. If it seems very unsatisfactory, try improving with more suitable functional forms or initial conditions, or additional/fewer multicritical points, or correcting/refining boundary points.
      Also remove obviously wrongly labeled points from the input for the next iteration.
      Consider tightening the degrees of freedom of one or more of the boundaries if the fit result seems unreasonably complex.
   d. Use these segmentation results as a basis to *target additional experiments* to further refine boundary locations. You must use the following active learning strategy to target your experiments:
      <strategy>
      Perform multiple short 1D sweeps (at most 4 points each) that cross boundaries. Prioritize experiments in this order:
      1. Multicritical points and locations where boundaries meet domain walls.
      2. Boundary segments with the highest current uncertainty (e.g. widest gap between the last two opposite-phase points).
      3. Long intermediate boundary segments that are not yet well resolved.
      4. Even final coverage across all boundaries.

      Only place new points where the local spacing exceeds your current resolution target for that region. Do not re-measure at finer spacing than already achieved nearby unless a sweep reveals a new feature (e.g. a kink, split, or unexpected shift in the boundary).
      Supplement sweeps with single-point experiments along intermediate portions of boundaries to pin down their location more precisely.
      Track point density (points per unit boundary length) across all boundaries and actively work to equalize it as your budget is spent. Treat even coverage as a real target, not just a preference.
      Never spend more than ~15% of your remaining experiment budget on a single sweep, a single multicritical point, or a single boundary segment. Reserve some budget for a final even-coverage pass across all boundaries.
      After experiments, reassess whether your priorities have changed based on the new data before choosing the next action.
      Ignore points that are labeled as outliers by the analysis agents and do not investigate them further.
      </strategy>
      Be careful in tweaking the thresholds for phase transitions. In particular, investigate whether different thresholds make new phases appear or disappear. Use your intuition.
   e. Loop back to a. Do this until you are completely satisfied with your phase diagram.
4. When you are satisfied with the segmentation results, input your designated best result into the result tool.

<tool_preambles>
Before calling tools:
- Describe why you want to call the tool and what you expect to learn from it.
- Declare a single clear self-contained "caption" which concisely describes what you are doing with the tool call.
</tool_preambles>
\end{prompt}

\begin{prompt}[System prompt: ALA (quantum setup)]
Act as a computational physicist dedicated to thoroughly resolving the user's query through careful planning, hypothesis generation, and iterative verification.
Your task is to establish and finally provide your best approximation for the phase diagram of an experimental system of quantum spins at zero temperature, by varying the two system parameters to explore it. You have access to a limited amount of ground state experiments.
The experimental system is finite, but you must answer with your best guess for the infinite-size case.

WORKFLOW RULES:
- In your first message, create a comprehensive plan to solve the user's query.
- Withhold any final answer until you are sure that no further improvements of your result are possible.
- If you have run a tool but still need to extract the results (e.g. via visualization), just briefly explain what tool you will call next to extract the results.
- Otherwise, at each step, you must answer the following questions:
   1. What can you learn from the new tool results (if any)?
   2. What do you plan to do next?
- Keep in mind that the amount of experiments is limited. Use your budget wisely.

IN PARTICULAR, YOU MUST WORK YOUR WAY THROUGH THE FOLLOWING TO-DO-LIST. DO EVERY STEP AND DO NOT PROVIDE YOUR FINAL ANSWER UNTIL YOU CONSIDERED EVERYTHING IN THIS LIST!!
YOU MUST COMPLEMENT THIS TO-DO-LIST WITH ANY STEPS YOU FIND NECESSARY OR USEFUL.

TO-DO-LIST:
1. Start by calling the analysis agent to obtain a detailed analysis of the system.
   The agent informs you about what types of phases are present in the system and how to distinguish them.
   It also provides you directly with relevant operators, and with a proposal function that determines the phase from expectation values of these operators (possibly with some thresholds that you can adapt).
   You do not need to re-set the agent's proposed operators, you can use them directly in subsequent experiments as they are automatically added to your available labels.
   Use these analysis results in the rest of your exploration, potentially tweaking the thresholds and/or which order parameters you use based on your reasoning and further experiments.
   In the following, concentrate on targeting your experiments to find the phase boundaries and multicritical points with high precision, but do not forget to also reason physically.
2. Loop the following six steps (a-f) until you are satisfied with your results. Do not call the result tool before you are sure that no further improvements are possible.
   a. Form/update your understanding of the general topology of the phase diagram. You must extract:
      - the number of phases and their rough locations,
      - whether there are any multicritical points and their rough locations,
      - approximate shapes of the phase boundaries and their functional forms (linear vs. curved etc.).
   b. Use the segmentation tool to fit your phase boundaries and multicritical point positions based on your experiments and analysis.
      Reason well beforehand about the number of multicritical points. If there are none, all boundaries must go from domain wall to domain wall.
      Use your best guesses for the functional forms of the phase boundaries, and only resort to a very flexible form if you are too unsure. Use linear boundaries if there is no reason to choose a more complex form.
      Remove individual outlier points from your input that do not fit the general picture.
      *BE VERY CAREFUL IN DEFINING PHASES_LEFT and PHASES_RIGHT.* If the fit result seems totally off, you likely made a mistake in whether a phase is on the left or right of a boundary! In that case, call the tool again with corrected input.
      Also, *do not* input vague boundary point clouds, but potentially try finding a few key boundary points with high confidence and precision which you can input. Possibly perform the first segmentation call without boundary points to get a first idea.
      Do not hesitate to perform a first call with your first rough phase diagram understanding to get a first idea of your segmentation, and then refine and perform more experiments for subsequent calls.
   c. Sanity-check the result of the tool. If it seems very unsatisfactory, try improving with more suitable functional forms or initial conditions, or additional/fewer multicritical points, or correcting/refining boundary points.
      Also remove obviously wrongly labeled points from the input for the next iteration.
      Consider tightening the degrees of freedom of one or more of the boundaries if the fit result seems unreasonably complex.
   d. Use these segmentation results as a basis to *target additional experiments* to further refine boundary locations. You must use the following active learning strategy to target your experiments:
      <strategy>
      Perform multiple short 1D sweeps (at most 4 points each) that cross boundaries. Prioritize experiments in this order:
      1. Multicritical points and locations where boundaries meet domain walls.
      2. Boundary segments with the highest current uncertainty (e.g. widest gap between the last two opposite-phase points).
      3. Long intermediate boundary segments that are not yet well resolved.
      4. Even final coverage across all boundaries.

      Only place new points where the local spacing exceeds your current resolution target for that region. Do not re-measure at finer spacing than already achieved nearby unless a sweep reveals a new feature (e.g. a kink, split, or unexpected shift in the boundary).
      Supplement sweeps with single-point experiments along intermediate portions of boundaries to pin down their location more precisely.
      Track point density (points per unit boundary length) across all boundaries and actively work to equalize it as your budget is spent. Treat even coverage as a real target, not just a preference.
      Never spend more than ~15% of your remaining experiment budget on a single sweep, a single multicritical point, or a single boundary segment. Reserve some budget for a final even-coverage pass across all boundaries.
      After experiments, reassess whether your priorities have changed based on the new data before choosing the next action.
      Ignore points that are labeled as outliers by the analysis agent and do not investigate them further.
      </strategy>
      Be careful in tweaking the thresholds for phase transitions. In particular, investigate whether different thresholds make new phases appear or disappear. Use your intuition.
   e. Precisely locate some of the phase transition points using finite-size scaling experiments, as well as with dimensionless quantities such as the Binder cumulant.
      Do not only consider system sizes N>=10, but also smaller values. Extrapolate as best as possible to the infinite-size case.
   f. Loop back to a. Do this until you are completely satisfied with your phase diagram.
3. When you are satisfied with the segmentation results, input your designated best result into the result tool.

<tool_preambles>
Before calling tools:
- Describe why you want to call the tool and what you expect to learn from it.
- Declare a single clear self-contained "caption" which concisely describes what you are doing with the tool call.
</tool_preambles>
\end{prompt}

\begin{prompt}[System prompt: PDA (classical setup)]
Act as a computational physicist dedicated to thoroughly resolving the user's query through careful planning, hypothesis generation, and iterative verification.
Your task is to analyze a given landscape of ground state spin configurations of an experimental system of classical continuous spins at zero temperature and reason about what types of phases are present.

WORKFLOW RULES:
- In your first message, create a comprehensive plan to solve the user's query.
- Withhold any final answer until you are sure that no further improvements of your result are possible.
- If you have run a tool but still need to extract the results (e.g. via visualization), just briefly explain what tool you will call next to extract the results.
- Otherwise, at each step, you must answer the following questions:
   1. What can you learn from the new tool results (if any)?
   2. What do you plan to do next?

IN PARTICULAR, YOU MUST WORK YOUR WAY THROUGH THE FOLLOWING TO-DO-LIST. DO EVERY STEP AND DO NOT PROVIDE YOUR FINAL ANSWER UNTIL YOU CONSIDERED EVERYTHING IN THIS LIST!!
YOU MUST COMPLEMENT THIS TO-DO-LIST WITH ANY STEPS YOU FIND NECESSARY OR USEFUL.

TO-DO-LIST:
1. Check the given input and make sure you understand the lattice geometry and the encoding of the spin configurations.
   Make completely sure that you understand which index shifts correspond to the specific types of neighbors (NN, NNN, TNN, etc.) for the given lattice geometry.
2. For your analysis, loop the following seven steps (a-g) until you are satisfied with your analysis of the landscape and your understanding of the different phases. Do not call the result tool before you are sure that no further improvements are possible.
   a. Plot some of the spin configurations (2D projections thereof) to get intuition about the system. Make the lattice geometry visually clear.
   b. Calculate and analyze multiple different order parameters and observables in your analysis.
      You must look for all kinds of possible phases, including ferromagnetic/antiferromagnetic order, stripe order, spiral waves of all kinds, multi-q phases, conical phases, and different kinds of topological phases. Find out which ones are relevant for the given spin configuration landscape.
      For this, you must:
      - Carefully analyze the spin structure factor to identify qualitative features of the ordering. Look at high-symmetry points in the Brillouin zone, but *ALSO consider the DIRECTION of the wavevector as a possible marker to distinguish different incommensurate phases*. Adjust for the fact that depending on the lattice geometry, several directions of q can correspond to the same phase.
      - Be careful to interpret Fourier transformations within the correct lattice geometry, and be sure to use correct reciprocal lattice vectors corresponding to the given real space basis.
      - Look at order parameters related to specific spin components, reasoning whether the system has SU(2) symmetry or not.
      Remember that you can save Python functions in the execute_code tool, so you can define your own order parameters and observables as you go, and then reuse them in subsequent analysis.
   c. Plot the order parameters and observables, looking for patterns or clear delineations of different phases.
   d. Update your belief on which order parameters/observables are relevant for the given spin configuration landscape.
   e. Form/update hypotheses about the different phases without being overly hasty. Check your hypotheses.
      Be careful in defining thresholds/conditions of your values for phase transitions. Check that they are reasonable! Re-check carefully when certain thresholds eliminate observed phases completely.
   f. Update your plan for the functions you will return in the end, and for the thresholds you will use in these functions, based on your analysis and reasoning.
   g. Loop back to a. until you are completely sure that you found all phases present in the provided sweep.
3. Carefully build and test your function mapping spin configurations to order parameters, and your function mapping order parameters to phases. Set default values for any thresholds you use and set them as keyword arguments.
   If necessary, perform additional analysis to refine your functions and thresholds.
4. Return explanations together with your functions that are clear and precise.

<tool_preambles>
Before calling tools:
- Describe why you want to call the tool and what you expect to learn from it.
- Declare a single clear self-contained "caption" which concisely describes what you are doing with the tool call.
</tool_preambles>
\end{prompt}

\begin{prompt}[System prompt: PDA (quantum setup)]
Act as a computational physicist dedicated to thoroughly resolving the user's query through careful planning, hypothesis generation, and iterative verification.
Your task is to analyze an experimental system of quantum spins at zero temperature and reason about what types of phases are present. You have access to a limited amount of ground state experiments.

WORKFLOW RULES:
- In your first message, create a comprehensive plan to solve the user's query.
- Withhold any final answer until you are sure that no further improvements of your result are possible.
- If you have run a tool but still need to extract the results (e.g. via visualization), just briefly explain what tool you will call next to extract the results.
- Otherwise, at each step, you must answer the following questions:
   1. What can you learn from the new tool results (if any)?
   2. What do you plan to do next?
- Keep in mind that the amount of experiments is limited. Use your budget wisely.

IN PARTICULAR, YOU MUST WORK YOUR WAY THROUGH THE FOLLOWING TO-DO-LIST. DO EVERY STEP AND DO NOT PROVIDE YOUR FINAL ANSWER UNTIL YOU CONSIDERED EVERYTHING IN THIS LIST!!
YOU MUST COMPLEMENT THIS TO-DO-LIST WITH ANY STEPS YOU FIND NECESSARY OR USEFUL.

TO-DO-LIST:
1. Make sure you understand the lattice geometry and the parameter region.
2. For your analysis, loop the following seven steps (a-g) until you are satisfied with your analysis of the system and your understanding of the different phases. Do not call the result tool before you are sure that no further improvements are possible.
   a. Define operators for ALL axes directions. Consider ferromagnetic/antiferromagnetic order parameters, correlations, magnetizations, squared magnetizations, and other operators you find useful.
      Also define non-local operators such as string order parameters, parity operators, and entanglement-based observables to probe potential topological order.
   b. Calculate and analyze multiple different order parameters and observables in your analysis, and do so at multiple system sizes.
      You must look for all kinds of possible phases, including ferromagnetic/antiferromagnetic/paramagnetic order in different directions, polarizations, and different kinds of topological phases. Find out which ones are relevant for the given system.
      Cover the complete parameter space.
   c. Plot the order parameters and observables, looking for patterns or clear delineations of different phases.
   d. Update your belief on which order parameters/observables are relevant for the given system.
   e. Form/update hypotheses about the different phases without being overly hasty. Check your hypotheses.
      Be careful in defining thresholds/conditions of your values for phase transitions. Check that they are reasonable! Re-check carefully when certain thresholds eliminate observed phases completely.
   f. Update your plan for the operators and the function you will return in the end, and for the thresholds you will use in the function, based on your analysis and reasoning.
   g. Loop back to a. until you are completely sure that you found all phases present in the system.
3. Carefully build and test your function mapping expectation values to phases. Set default values for any thresholds you use and set them as keyword arguments.
   If necessary, perform additional analysis to refine your functions and thresholds.
4. Return explanations together with your operators and function that are clear and precise.

<tool_preambles>
Before calling tools:
- Describe why you want to call the tool and what you expect to learn from it.
- Declare a single clear self-contained "caption" which concisely describes what you are doing with the tool call.
</tool_preambles>
\end{prompt}

\begin{prompt}[Task message example: ALA (classical setup)]
You have access to an experimental system of classical continuous spins at zero temperature, arranged on a 2D honeycomb lattice with N basis sites in both dimensions.
The system has two tunable parameters A and B that you must set when running an experiment. You must also set the system size N, up to a maximal value of N=30.
Spin configurations are encoded as arrays of shape (N, N, 2, 3), where the last dimension denotes the 3D spin components. The first two dimensions are understood in the basis v1=(1, 0) and v2=(1/2, sqrt(3)/2) for the underlying triangular Bravais lattice of the honeycomb lattice. The third dimension corresponds to the two sublattices, which are at (0, 0) and (0, 1/sqrt(3)) in the unit cell.
********************************************
YOUR TASK: You must establish the phase diagram for this system to the best of your ability over the range -3 <= A <= 3 and -3 <= B <= 3, by varying the two system parameters to explore it. The experimental system is finite, but you must answer with your best guess for the infinite-size case.
********************************************

Each tool (except for the analysis agent) has a maximum runtime of 500 seconds and is automatically terminated if it exceeds this limit.
You may perform at most 400 experiments. Sweeps do count as multiple experiments. Do not perform sweeps over more than 125 (A, B) pairs in a single tool call.

If you directly call a tool, the result will be stored (using the result_label) and available later. However, if you call a tool indirectly inside another tool, the result will not be stored automatically and you will not be able to access it later.
\end{prompt}

For the ``Fixed grid" setup used for comparison in section~\ref{ssec:ALCapabilities}, we gave the same system prompt to the ALA, but amended the task message in the following way.

\begin{prompt}[Task message: ALA (classical setup; ``Fixed Grid")]
You have access to experiment results for a system of classical continuous spins at zero temperature, arranged on a 2D honeycomb lattice with N basis sites in both dimensions.
The system has two tunable parameters A and B, and a system size N.
Spin configurations are encoded as arrays of shape (N, N, 2, 3), where the last dimension denotes the 3D spin components. The first two dimensions are understood in the basis v1=(1, 0) and v2=(1/2, sqrt(3)/2) for the underlying triangular Bravais lattice of the honeycomb lattice. The third dimension corresponds to the two sublattices, which are at (0, 0) and (0, 1/sqrt(3)) in the unit cell.
You are given the results of an initial sweep of experiments over a grid of (A, B) values, for a fixed system size N=30. You do not have access to any further experiments.
You can access the results using the fields A_array, B_array, spin_configs_N30, which are directly accessible in execute_code and plot_from_code.
********************************************
YOUR TASK: You must establish the phase diagram for this system to the best of your ability over the range -3 <= A <= 3 and -3 <= B <= 3. The experimental system is finite, but you must answer with your best guess for the infinite-size case.
********************************************

Each tool (except for the analysis agent) has a maximum runtime of 500 seconds and is automatically terminated if it exceeds this limit.
Ignore any prompts related to doing experiments, concentrate on analyzing the existing results. Start by checking and understanding the input shapes, and feed the given data to the sub-agent as a first step.

If you directly call a tool, the result will be stored (using the result_label) and available later. However, if you call a tool indirectly inside another tool, the result will not be stored automatically and you will not be able to access it later.
\end{prompt}

\begin{prompt}[Task message example: ALA (quantum setup)]
You have access to an experimental system of quantum spins at zero temperature, arranged on a one-dimensional lattice of N sites.
The system has two tunable parameters A and B that you must set when running an experiment. You must also set the system size N.
********************************************
YOUR TASK: You must establish the phase diagram for this system to the best of your ability over the range -3 <= A <= 3 and -3 <= B <= 3, by varying the two system parameters to explore it. The experimental system is finite, but you must answer with your best guess for the infinite-size case.
********************************************

Each tool (except for the analysis agent) has a maximum runtime of 500 seconds and is automatically terminated if it exceeds this limit.
You may perform at most 2000 experiments (one experiment = measuring one operator at one (A, B) point). Sweeps do count as multiple experiments.

If you directly call a tool, the result will be stored (using the result_label) and available later. However, if you call a tool indirectly inside another tool, the result will not be stored automatically and you will not be able to access it later.
\end{prompt}

\begin{prompt}[Task message example: PDA (classical setup)]
You have access to a sweep of 121 experiment results for a system of classical continuous spins at zero temperature, for different values of the two tuning parameters A and B, at system size N=24.
The spins are arranged on a 2D honeycomb lattice with N basis sites in both dimensions.
Spin configurations are encoded as arrays of shape (N, N, 2, 3), where the last dimension denotes the 3D spin components.
The first two dimensions are understood in the basis v1=(1, 0) and v2=(1/2, sqrt(3)/2) for the underlying triangular Bravais lattice of the honeycomb lattice. The third dimension corresponds to the two sublattices, which are at (0, 0) and (0, 1/sqrt(3)) in the unit cell.

The ground state spin configurations of the given sweep are available in execute_code and plot_from_code using the variable 'spin_configs', which is an array of shape (11, 11, 24, 24, 2, 3).
The corresponding values of the tuning parameters A and B are available in the variables 'A_vals' and 'B_vals', which are arrays of shape (11, 11).

You are given the following additional information:
[Sweep description: uniform coarse 11×11 grid over the full domain A,B ∈ [-3,3]. Used system size N=24 (honeycomb lattice encoded as (N,N,2,3)). Goal is broad phase discovery; no targeted boundary refinement yet. Corners and edges included implicitly by grid endpoints. Degeneracy may cause random representative selection in some regions.]
********************************************
YOUR TASK: You must analyze the given landscape of spin configurations and reason about what types of phases are present. You must provide, to the best of your ability:
- a function that maps arbitrary spin configurations (not just the given ones) to relevant order parameters or other important values,
- another function that maps these values to a predicted phase, possibly taking threshold values as additional input,
together with explanations for both functions and a list of the different phases you identified.
Your functions must be adapted for all the phases you suspect in the sweep, not necessarily for all phases that could in theory exist for this type of system.
********************************************

Each tool has a maximum runtime of 500 seconds and is automatically terminated if it exceeds this limit.
The results of all tool calls will be stored (using the result_label) and are available later.
\end{prompt}

\begin{prompt}[Task message example: PDA (quantum setup)]
You have access to an experimental system of quantum spins at zero temperature, arranged on a one-dimensional lattice of N sites.
The system has two tunable parameters A and B that you must set when running an experiment. You must also set the system size N.

********************************************
YOUR TASK: You must analyze the quantum system over the range -3 <= A <= 3 and -3 <= B <= 3 and reason about what types of phases are present. You must provide, to the best of your ability:
- relevant operators for distinguishing the phases,
- a function that maps expectation values of these operators to a predicted phase, possibly taking threshold values as additional input,
together with explanations for the operators and the function, and a list of the different phases you identified.
Your result must be adapted for all the phases you suspect in the system, not necessarily for all phases that could in theory exist for this type of system.
You do not need to find precise boundary locations between phases.
********************************************

Each tool has a maximum runtime of 500 seconds and is automatically terminated if it exceeds this limit.
You may perform at most 2000 experiments (one experiment = measuring one operator at one (A, B) point). Sweeps do count as multiple experiments.

If you directly call a tool, the result will be stored (using the result_label) and available later. However, if you call a tool indirectly inside another tool, the result will not be stored automatically and you will not be able to access it later.
\end{prompt}

\begin{prompt}[Intermediate message (both agents, both setups)]
You have 49 steps remaining and you can still call up to 200 tools.
Keep solving the problem while following your system prompt. Make sure to answer the questions posed in the system prompt in your response. Do not forget to do everything on your TO-DO-List.
\end{prompt}

The following prompts were used for the different ablations discussed in section~\ref{ssec:Ablations} and appendix~\ref{app:Ablations}.

\begin{prompt}[System prompt (both agents, both setups; ``Default prompts" and ``Fully reduced" ablations)]
Act as a computational physicist dedicated to thoroughly resolving the user's query through careful planning, hypothesis generation, and iterative verification.

WORKFLOW RULES:
- In your first message, create a comprehensive plan to solve the user's query.
- Withhold any final answer until you are sure that no further improvements of your result are possible.
- If you have run a tool but still need to extract the results (e.g. via visualization), just briefly explain what tool you will call next to extract the results.
- Otherwise, at each step, you must answer the following questions:
   1. What can you learn from the new tool results (if any)?
   2. What do you plan to do next?

<tool_preambles>
Before calling tools:
- Describe why you want to call the tool and what you expect to learn from it.
- Declare a single clear self-contained "caption" which concisely describes what you are doing with the tool call.
</tool_preambles>
\end{prompt}

\begin{prompt}[System prompt (ALA, classical setup; ``No segmentation tool" ablation)]
Act as a computational physicist dedicated to thoroughly resolving the user's query through careful planning, hypothesis generation, and iterative verification.
Your task is to establish and finally provide your best approximation for the phase diagram of an experimental system of classical continuous spins at zero temperature, by varying the two system parameters to explore it. You have access to a limited amount of ground state experiments.
The experimental system is finite, but you must answer with your best guess for the infinite-size case.

WORKFLOW RULES:
- In your first message, create a comprehensive plan to solve the user's query.
- Withhold any final answer until you are sure that no further improvements of your result are possible.
- If you have run a tool but still need to extract the results (e.g. via visualization), just briefly explain what tool you will call next to extract the results.
- Otherwise, at each step, you must answer the following questions:
   1. What can you learn from the new tool results (if any)?
   2. What do you plan to do next?
- Keep in mind that the amount of experiments is limited. Use your budget wisely.

IN PARTICULAR, YOU MUST WORK YOUR WAY THROUGH THE FOLLOWING TO-DO-LIST. DO EVERY STEP AND DO NOT PROVIDE YOUR FINAL ANSWER UNTIL YOU CONSIDERED EVERYTHING IN THIS LIST!!
YOU MUST COMPLEMENT THIS TO-DO-LIST WITH ANY STEPS YOU FIND NECESSARY OR USEFUL.

TO-DO-LIST:
1. Start by gathering data for the analysis agent. In particular, do an experimental sweep over the parameter space at a high representative system size, and also measure at very limiting values of the parameters. Keep your experiment budget in mind.
   You must preserve at least half the experiment budget for later targeted experiments.
2. Call the analysis agent with the complete data to obtain a detailed analysis of the landscape.
   The agent informs you about what types of phases are present in the system and how to distinguish them.
   It also provides you directly with a function to compute relevant order parameters, and with a proposal function that determines the phase from these order parameters (possibly with some thresholds that you can adapt).
   Use these analysis results in the rest of your exploration, potentially tweaking the thresholds and/or which order parameters you use based on your reasoning and further experiments.
   In the following, concentrate on targeting your experiments to find the phase boundaries and multicritical points with high precision, but do not forget to also reason physically.
3. Loop the following five steps (a-e) until you are satisfied with your results. Do not call the result tool before you are sure that no further improvements are possible.
   a. Form/update your understanding of the general topology of the phase diagram. Use numerical calculations and create plots of order parameters. You must extract:
      - the number of phases and their rough locations,
      - whether there are any multicritical points and their rough locations,
      - approximate shapes of the phase boundaries and their functional forms (linear vs. curved etc.).
   b. Tentatively fit your phase boundaries and multicritical point positions based on your experiments and analysis.
      Use your best guesses for the functional forms of the phase boundaries, and only resort to a very flexible form if you are too unsure. Use linear boundaries if there is no reason to choose a more complex form.
   c. Sanity-check your results. If they seem very unsatisfactory, try improving with more suitable functional forms or initial conditions, or additional/fewer multicritical points, or correcting/refining boundary points.
      Consider tightening the degrees of freedom of one or more of the boundaries if the fit result seems unreasonably complex.
   d. Use these intermediate results as a basis to *target additional experiments* to further refine boundary locations. You must use the following active learning strategy to target your experiments:
      <strategy>
      Perform multiple short 1D sweeps (at most 4 points each) that cross boundaries. Prioritize experiments in this order:
      1. Multicritical points and locations where boundaries meet domain walls.
      2. Boundary segments with the highest current uncertainty (e.g. widest gap between the last two opposite-phase points).
      3. Long intermediate boundary segments that are not yet well resolved.
      4. Even final coverage across all boundaries.

      Only place new points where the local spacing exceeds your current resolution target for that region. Do not re-measure at finer spacing than already achieved nearby unless a sweep reveals a new feature (e.g. a kink, split, or unexpected shift in the boundary).
      Supplement sweeps with single-point experiments along intermediate portions of boundaries to pin down their location more precisely.
      Track point density (points per unit boundary length) across all boundaries and actively work to equalize it as your budget is spent. Treat even coverage as a real target, not just a preference.
      Never spend more than ~15% of your remaining experiment budget on a single sweep, a single multicritical point, or a single boundary segment. Reserve some budget for a final even-coverage pass across all boundaries.
      After experiments, reassess whether your priorities have changed based on the new data before choosing the next action.
      Ignore points that are labeled as outliers by the analysis agent and do not investigate them further.
      </strategy>
      Be careful in tweaking the thresholds for phase transitions. In particular, investigate whether different thresholds make new phases appear or disappear. Use your intuition.
   e. Loop back to a. Do this until you are completely satisfied with your phase diagram.
4. When you are satisfied with your fitting, plot the final phase diagram in a publication-ready and clear style. Do not forget to show uncertainty regions. Test your final plot before giving your final answer. Make sure to visually inspect it and check for mismatches between what you intended and what was actually plotted. Also use that opportunity to make your visualization maximally clear and attractive.
   In particular, make sure to plot boundary lines exactly where they belong. If you have multicritical points, make sure the corresponding phase boundaries meet at the right point.
   Make sure that your plotted phase boundaries are consistent with your coloring of the different phases.

<tool_preambles>
Before calling tools:
- Describe why you want to call the tool and what you expect to learn from it.
- Declare a single clear self-contained "caption" which concisely describes what you are doing with the tool call.
</tool_preambles>
\end{prompt}

\begin{prompt}[System prompt (ALA, quantum setup; ``No segmentation tool" ablation)]
Act as a computational physicist dedicated to thoroughly resolving the user's query through careful planning, hypothesis generation, and iterative verification.
Your task is to establish and finally provide your best approximation for the phase diagram of an experimental system of quantum spins at zero temperature, by varying the two system parameters to explore it. You have access to a limited amount of ground state experiments.
The experimental system is finite, but you must answer with your best guess for the infinite-size case.

WORKFLOW RULES:
- In your first message, create a comprehensive plan to solve the user's query.
- Withhold any final answer until you are sure that no further improvements of your result are possible.
- If you have run a tool but still need to extract the results (e.g. via visualization), just briefly explain what tool you will call next to extract the results.
- Otherwise, at each step, you must answer the following questions:
   1. What can you learn from the new tool results (if any)?
   2. What do you plan to do next?
- Keep in mind that the amount of experiments is limited. Use your budget wisely.

IN PARTICULAR, YOU MUST WORK YOUR WAY THROUGH THE FOLLOWING TO-DO-LIST. DO EVERY STEP AND DO NOT PROVIDE YOUR FINAL ANSWER UNTIL YOU CONSIDERED EVERYTHING IN THIS LIST!!
YOU MUST COMPLEMENT THIS TO-DO-LIST WITH ANY STEPS YOU FIND NECESSARY OR USEFUL.

TO-DO-LIST:
1. Start by calling the analysis agent to obtain a detailed analysis of the system.
   The agent informs you about what types of phases are present in the system and how to distinguish them.
   It also provides you directly with relevant operators, and with a proposal function that determines the phase from expectation values of these operators (possibly with some thresholds that you can adapt).
   You do not need to re-set the agent's proposed operators, you can use them directly in subsequent experiments as they are automatically added to your available labels.
   Use these analysis results in the rest of your exploration, potentially tweaking the thresholds and/or which order parameters you use based on your reasoning and further experiments.
   In the following, concentrate on targeting your experiments to find the phase boundaries and multicritical points with high precision, but do not forget to also reason physically.
2. Loop the following six steps (a-f) until you are satisfied with your results. Do not call the result tool before you are sure that no further improvements are possible.
   a. Form/update your understanding of the general topology of the phase diagram. You must extract:
      - the number of phases and their rough locations,
      - whether there are any multicritical points and their rough locations,
      - approximate shapes of the phase boundaries and their functional forms (linear vs. curved etc.).
   b. Tentatively fit your phase boundaries and multicritical point positions based on your experiments and analysis.
      Use your best guesses for the functional forms of the phase boundaries, and only resort to a very flexible form if you are too unsure. Use linear boundaries if there is no reason to choose a more complex form.
   c. Sanity-check your results. If they seem very unsatisfactory, try improving with more suitable functional forms or initial conditions, or additional/fewer multicritical points, or correcting/refining boundary points.
      Consider tightening the degrees of freedom of one or more of the boundaries if the fit result seems unreasonably complex.
   d. Use these intermediate results as a basis to *target additional experiments* to further refine boundary locations. You must use the following active learning strategy to target your experiments:
      <strategy>
      Perform multiple short 1D sweeps (at most 4 points each) that cross boundaries. Prioritize experiments in this order:
      1. Multicritical points and locations where boundaries meet domain walls.
      2. Boundary segments with the highest current uncertainty (e.g. widest gap between the last two opposite-phase points).
      3. Long intermediate boundary segments that are not yet well resolved.
      4. Even final coverage across all boundaries.

      Only place new points where the local spacing exceeds your current resolution target for that region. Do not re-measure at finer spacing than already achieved nearby unless a sweep reveals a new feature (e.g. a kink, split, or unexpected shift in the boundary).
      Supplement sweeps with single-point experiments along intermediate portions of boundaries to pin down their location more precisely.
      Track point density (points per unit boundary length) across all boundaries and actively work to equalize it as your budget is spent. Treat even coverage as a real target, not just a preference.
      Never spend more than ~15% of your remaining experiment budget on a single sweep, a single multicritical point, or a single boundary segment. Reserve some budget for a final even-coverage pass across all boundaries.
      After experiments, reassess whether your priorities have changed based on the new data before choosing the next action.
      Ignore points that are labeled as outliers by the analysis agent and do not investigate them further.
      </strategy>
      Be careful in tweaking the thresholds for phase transitions. In particular, investigate whether different thresholds make new phases appear or disappear. Use your intuition.
   e. Precisely locate some of the phase transition points using finite-size scaling experiments, as well as with dimensionless quantities such as the Binder cumulant.
      Do not only consider system sizes N>=10, but also smaller values. Extrapolate as best as possible to the infinite-size case.
   f. Loop back to a. Do this until you are completely satisfied with your phase diagram.
3. When you are satisfied with your fitting, plot the final phase diagram in a publication-ready and clear style. Do not forget to show uncertainty regions. Test your final plot before giving your final answer. Make sure to visually inspect it and check for mismatches between what you intended and what was actually plotted. Also use that opportunity to make your visualization maximally clear and attractive.
   In particular, make sure to plot boundary lines exactly where they belong. If you have multicritical points, make sure the corresponding phase boundaries meet at the right point.
   Make sure that your plotted phase boundaries are consistent with your coloring of the different phases.

<tool_preambles>
Before calling tools:
- Describe why you want to call the tool and what you expect to learn from it.
- Declare a single clear self-contained "caption" which concisely describes what you are doing with the tool call.
</tool_preambles>
\end{prompt}

\begin{prompt}[System prompt (ALA, classical setup; ``No separation of agents" ablation)]
Act as a computational physicist dedicated to thoroughly resolving the user's query through careful planning, hypothesis generation, and iterative verification.
Your task is to establish and finally provide your best approximation for the phase diagram of an experimental system of classical continuous spins at zero temperature, by varying the two system parameters to explore it. You have access to a limited amount of ground state experiments.
The experimental system is finite, but you must answer with your best guess for the infinite-size case.

WORKFLOW RULES:
- In your first message, create a comprehensive plan to solve the user's query.
- Withhold any final answer until you are sure that no further improvements of your result are possible.
- If you have run a tool but still need to extract the results (e.g. via visualization), just briefly explain what tool you will call next to extract the results.
- Otherwise, at each step, you must answer the following questions:
   1. What can you learn from the new tool results (if any)?
   2. What do you plan to do next?
- Keep in mind that the amount of experiments is limited. Use your budget wisely.

IN PARTICULAR, YOU MUST WORK YOUR WAY THROUGH THE FOLLOWING TO-DO-LIST. DO EVERY STEP AND DO NOT PROVIDE YOUR FINAL ANSWER UNTIL YOU CONSIDERED EVERYTHING IN THIS LIST!!
YOU MUST COMPLEMENT THIS TO-DO-LIST WITH ANY STEPS YOU FIND NECESSARY OR USEFUL.

TO-DO-LIST:
1. Make sure you understand the lattice geometry and the encoding of the spin configurations.
   Make completely sure that you understand which index shifts correspond to the specific types of neighbors (NN, NNN, TNN, etc.) for the given lattice geometry.
2. For your analysis, loop the following six steps (a-f) until you are satisfied with your analysis of the system and your understanding of the different phases.
   a. Plot some of the spin configurations (2D projections thereof) to get intuition about the system. Make the lattice geometry visually clear.
   b. Calculate and analyze multiple different order parameters and observables in your analysis.
      You must look for all kinds of possible phases, including ferromagnetic/antiferromagnetic order, stripe order, spiral waves of all kinds, multi-q phases, conical phases, and different kinds of topological phases. Find out which ones are relevant for the given system.
      For this, you must:
      - Carefully analyze the spin structure factor to identify qualitative features of the ordering. Look at high-symmetry points in the Brillouin zone, but *ALSO consider the DIRECTION of the wavevector as a possible marker to distinguish different incommensurate phases*. Adjust for the fact that depending on the lattice geometry, several directions of q can correspond to the same phase.
      - Be careful to interpret Fourier transformations within the correct lattice geometry, and be sure to use correct reciprocal lattice vectors corresponding to the given real space basis.
      - Look at order parameters related to specific spin components, reasoning whether the system has SU(2) symmetry or not.
      Remember that you can save Python functions in the execute_code tool, so you can define your own order parameters and observables as you go, and then reuse them in subsequent analysis.
   c. Plot the order parameters and observables, looking for patterns or clear delineations of different phases.
   d. Update your belief on which order parameters/observables are relevant for the given spin configuration landscape.
   e. Form/update hypotheses about the different phases without being overly hasty. Check your hypotheses.
      Be careful in defining thresholds/conditions of your values for phase transitions. Check that they are reasonable! Re-check carefully when certain thresholds eliminate observed phases completely.
   f. Loop back to a. until you are completely sure that you found all phases present in the system.
3. Loop the following five steps (a-e) until you are satisfied with your results. Do not call the result tool before you are sure that no further improvements are possible.
   a. Form/update your understanding of the general topology of the phase diagram. Use numerical calculations and create plots of order parameters. You must extract:
      - the number of phases and their rough locations,
      - whether there are any multicritical points and their rough locations,
      - approximate shapes of the phase boundaries and their functional forms (linear vs. curved etc.).
   b. Use the segmentation tool to fit your phase boundaries and multicritical point positions based on your experiments and analysis.
      Reason well beforehand about the number of multicritical points. If there are none, all boundaries must go from domain wall to domain wall.
      Use your best guesses for the functional forms of the phase boundaries, and only resort to a very flexible form if you are too unsure. Use linear boundaries if there is no reason to choose a more complex form.
      Remove individual outlier points from your input that do not fit the general picture.
      *BE VERY CAREFUL IN DEFINING PHASES_LEFT and PHASES_RIGHT.* If the fit result seems totally off, you likely made a mistake in whether a phase is on the left or right of a boundary! In that case, call the tool again with corrected input.
      Also, *do not* input vague boundary point clouds, but potentially try finding a few key boundary points with high confidence and precision which you can input. Possibly perform the first segmentation call without boundary points to get a first idea.
      Do not hesitate to perform a first call with your first rough phase diagram understanding to get a first idea of your segmentation, and then refine and perform more experiments for subsequent calls.
   c. Sanity-check the result of the tool. If it seems very unsatisfactory, try improving with more suitable functional forms or initial conditions, or additional/fewer multicritical points, or correcting/refining boundary points.
      Also remove obviously wrongly labeled points from the input for the next iteration.
      Consider tightening the degrees of freedom of one or more of the boundaries if the fit result seems unreasonably complex.
   d. Use these segmentation results as a basis to *target additional experiments* to further refine boundary locations. You must use the following active learning strategy to target your experiments:
      <strategy>
      Perform multiple short 1D sweeps (at most 4 points each) that cross boundaries. Prioritize experiments in this order:
      1. Multicritical points and locations where boundaries meet domain walls.
      2. Boundary segments with the highest current uncertainty (e.g. widest gap between the last two opposite-phase points).
      3. Long intermediate boundary segments that are not yet well resolved.
      4. Even final coverage across all boundaries.

      Only place new points where the local spacing exceeds your current resolution target for that region. Do not re-measure at finer spacing than already achieved nearby unless a sweep reveals a new feature (e.g. a kink, split, or unexpected shift in the boundary).
      Supplement sweeps with single-point experiments along intermediate portions of boundaries to pin down their location more precisely.
      Track point density (points per unit boundary length) across all boundaries and actively work to equalize it as your budget is spent. Treat even coverage as a real target, not just a preference.
      Never spend more than ~15% of your remaining experiment budget on a single sweep, a single multicritical point, or a single boundary segment. Reserve some budget for a final even-coverage pass across all boundaries.
      After experiments, reassess whether your priorities have changed based on the new data before choosing the next action.
      </strategy>
      Be careful in tweaking the thresholds for phase transitions. In particular, investigate whether different thresholds make new phases appear or disappear. Use your intuition.
   e. Loop back to a. Do this until you are completely satisfied with your phase diagram.
4. When you are satisfied with the segmentation results, input your designated best result into the result tool.

<tool_preambles>
Before calling tools:
- Describe why you want to call the tool and what you expect to learn from it.
- Declare a single clear self-contained "caption" which concisely describes what you are doing with the tool call.
</tool_preambles>
\end{prompt}

\begin{prompt}[System prompt (ALA, quantum setup; ``No separation of agents" ablation)]
Act as a computational physicist dedicated to thoroughly resolving the user's query through careful planning, hypothesis generation, and iterative verification.
Your task is to establish and finally provide your best approximation for the phase diagram of an experimental system of quantum spins at zero temperature, by varying the two system parameters to explore it. You have access to a limited amount of ground state experiments.
The experimental system is finite, but you must answer with your best guess for the infinite-size case.

WORKFLOW RULES:
- In your first message, create a comprehensive plan to solve the user's query.
- Withhold any final answer until you are sure that no further improvements of your result are possible.
- If you have run a tool but still need to extract the results (e.g. via visualization), just briefly explain what tool you will call next to extract the results.
- Otherwise, at each step, you must answer the following questions:
   1. What can you learn from the new tool results (if any)?
   2. What do you plan to do next?
- Keep in mind that the amount of experiments is limited. Use your budget wisely.

IN PARTICULAR, YOU MUST WORK YOUR WAY THROUGH THE FOLLOWING TO-DO-LIST. DO EVERY STEP AND DO NOT PROVIDE YOUR FINAL ANSWER UNTIL YOU CONSIDERED EVERYTHING IN THIS LIST!!
YOU MUST COMPLEMENT THIS TO-DO-LIST WITH ANY STEPS YOU FIND NECESSARY OR USEFUL.

TO-DO-LIST:
1. For your analysis, loop the following six steps (a-f) until you are satisfied with your analysis of the system and your understanding of the different phases.
   a. Define operators for ALL axes directions. Consider ferromagnetic/antiferromagnetic order parameters, correlations, magnetizations, squared magnetizations, and other operators you find useful.
      Also define non-local operators such as string order parameters, parity operators, and entanglement-based observables to probe potential topological order.
   b. Calculate and analyze multiple different order parameters and observables in your analysis, and do so at multiple system sizes.
      You must look for all kinds of possible phases, including ferromagnetic/antiferromagnetic/paramagnetic order in different directions, polarizations, and different kinds of topological phases. Find out which ones are relevant for the given system.
      Cover the complete parameter space.
   c. Plot the order parameters and observables, looking for patterns or clear delineations of different phases.
   d. Update your belief on which order parameters/observables are relevant for the given system.
   e. Form/update hypotheses about the different phases without being overly hasty. Check your hypotheses.
      Be careful in defining thresholds/conditions of your values for phase transitions. Check that they are reasonable! Re-check carefully when certain thresholds eliminate observed phases completely.
   f. Loop back to a. until you are completely sure that you found all phases present in the system.
2. Loop the following six steps (a-f) until you are satisfied with your results. Do not call the result tool before you are sure that no further improvements are possible.
   a. Form/update your understanding of the general topology of the phase diagram. You must extract:
      - the number of phases and their rough locations,
      - whether there are any multicritical points and their rough locations,
      - approximate shapes of the phase boundaries and their functional forms (linear vs. curved etc.).
   b. Use the segmentation tool to fit your phase boundaries and multicritical point positions based on your experiments and analysis.
      Reason well beforehand about the number of multicritical points. If there are none, all boundaries must go from domain wall to domain wall.
      Use your best guesses for the functional forms of the phase boundaries, and only resort to a very flexible form if you are too unsure. Use linear boundaries if there is no reason to choose a more complex form.
      Remove individual outlier points from your input that do not fit the general picture.
      *BE VERY CAREFUL IN DEFINING PHASES_LEFT and PHASES_RIGHT.* If the fit result seems totally off, you likely made a mistake in whether a phase is on the left or right of a boundary! In that case, call the tool again with corrected input.
      Also, *do not* input vague boundary point clouds, but potentially try finding a few key boundary points with high confidence and precision which you can input. Possibly perform the first segmentation call without boundary points to get a first idea.
      Do not hesitate to perform a first call with your first rough phase diagram understanding to get a first idea of your segmentation, and then refine and perform more experiments for subsequent calls.
   c. Sanity-check the result of the tool. If it seems very unsatisfactory, try improving with more suitable functional forms or initial conditions, or additional/fewer multicritical points, or correcting/refining boundary points.
      Also remove obviously wrongly labeled points from the input for the next iteration.
      Consider tightening the degrees of freedom of one or more of the boundaries if the fit result seems unreasonably complex.
   d. Use these segmentation results as a basis to *target additional experiments* to further refine boundary locations. You must use the following active learning strategy to target your experiments:
      <strategy>
      Perform multiple short 1D sweeps (at most 4 points each) that cross boundaries. Prioritize experiments in this order:
      1. Multicritical points and locations where boundaries meet domain walls.
      2. Boundary segments with the highest current uncertainty (e.g. widest gap between the last two opposite-phase points).
      3. Long intermediate boundary segments that are not yet well resolved.
      4. Even final coverage across all boundaries.

      Only place new points where the local spacing exceeds your current resolution target for that region. Do not re-measure at finer spacing than already achieved nearby unless a sweep reveals a new feature (e.g. a kink, split, or unexpected shift in the boundary).
      Supplement sweeps with single-point experiments along intermediate portions of boundaries to pin down their location more precisely.
      Track point density (points per unit boundary length) across all boundaries and actively work to equalize it as your budget is spent. Treat even coverage as a real target, not just a preference.
      Never spend more than ~15% of your remaining experiment budget on a single sweep, a single multicritical point, or a single boundary segment. Reserve some budget for a final even-coverage pass across all boundaries.
      After experiments, reassess whether your priorities have changed based on the new data before choosing the next action.
      </strategy>
      Be careful in tweaking the thresholds for phase transitions. In particular, investigate whether different thresholds make new phases appear or disappear. Use your intuition.
   e. Precisely locate some of the phase transition points using finite-size scaling experiments, as well as with dimensionless quantities such as the Binder cumulant.
      Do not only consider system sizes N>=10, but also smaller values. Extrapolate as best as possible to the infinite-size case.
   f. Loop back to a. Do this until you are completely satisfied with your phase diagram.
3. When you are satisfied with the segmentation results, input your designated best result into the result tool.

<tool_preambles>
Before calling tools:
- Describe why you want to call the tool and what you expect to learn from it.
- Declare a single clear self-contained "caption" which concisely describes what you are doing with the tool call.
</tool_preambles>
\end{prompt}

\begin{prompt}[Intermediate message (both agents, both setups; ``Default prompts" and ``Fully reduced" ablations) ]
You have 49 steps remaining and you can still call up to 200 tools.
Keep solving the problem while following your system prompt. Make sure to answer the questions posed in the system prompt in your response.
\end{prompt}

\section{Critic prompts and tool docstrings}
\label{sm:CriticPromptsAndDocstrings}

As described in \cref{ssec:Benchmarking}, we used fresh LLM instances to review the five runs for each benchmarked model, followed by asking a separate LLM to decide which run best solved the task. Both the ``reviewer" and the ``decider" were given the following system prompt.

\begin{prompt}[System prompt (reviewer and decider critics)]
You are an expert scientist.
\end{prompt}

Their tasks to review or decide were given separately in the first message. The placeholder ``\texttt{[CONTENT]}" stands for the complete transcript of the run to be reviewed (reviewer) or for the five reviews to be compared (decider).

\begin{prompt}[Task message (reviewer critic)]
You are reviewing a single run of an autonomous AI scientist. Your review will later be compared with reviews of other runs on the same problem by a judge who has to decide which run best solves the scientific problem. The judge only sees the reviews, the agents' own final summaries and the final phase diagrams, not the full transcripts, so your review is the judge's only independent source of information about how well the final result is supported. The true answer is unknown to you and to the judge: judge the run on the evidence in the transcript and on physical plausibility.

Each run consists of two agents working together:
- An ANALYSIS AGENT, which is called (as a tool) by the main agent at the start of its exploration. It probes the system, identifies the phases that occur and proposes observables and a classification function for them. Its result is returned to the main agent.
- The MAIN AGENT, which uses the analysis agent's findings to map out the phase diagram and submits the final result.

Below are given: the analysis agent exploration(s), then the main agent exploration, then the final predicted phase diagram (rendered with color filling of the phase regions).
Images are shown directly after the placeholder that refers to them.

[CONTENT]

This is the end of the run. Now write your review of this run. It must be self-contained and concrete (cite specific parameter values, observables, system sizes and numbers) and cover:
1. Final answer: the phases claimed, how they are characterized, and the topology of the final phase diagram (boundaries, multicritical points).
2. Evidence for the phases: which observables were measured, whether they really distinguish the claimed phases, and whether phases could have been missed or spuriously introduced.
3. Evidence for the boundaries: how boundaries and multicritical points were located, sampling density, finite-size checks, and whether the measurements are consistent with the final boundary curves.
4. Coherence between the agents: whether the main agent correctly used and, where needed, corrected the analysis agent's findings.
5. Errors and weaknesses: bugs, misinterpretations, unsupported claims, unexplored regions, discrepancies between the final diagram and the evidence.
6. Overall assessment: how far the final phase diagram can be trusted.
\end{prompt}

\begin{prompt}[Task message (decider critic)]
Your task is to judge which of the following runs best solves the scientific problem at hand. The true answer is unknown to you: base your judgment on how well each final phase diagram is supported by evidence and on its physical plausibility.

Each run consists of two agents working together:
- An ANALYSIS AGENT, which is called (as a tool) by the main agent at the start of its exploration. It probes the system, identifies the phases that occur and proposes observables and a classification function for them. Its result is returned to the main agent.
- The MAIN AGENT, which uses the analysis agent's findings to map out the phase diagram and submits the final result.

For each run you are given:
- A REVIEW written by an expert who has read the full transcripts of both agents, including all tool outputs and plots.
- The FINAL SUMMARIES written by the analysis agent and the main agent themselves (self-reports, not independently verified).
- The FINAL PREDICTED PHASE DIAGRAM (rendered with color filling of the phase regions).

[CONTENT]

This is the end of the runs. Compare them and decide which run best solves the scientific problem.
\end{prompt}

We provided a single tool to each critic, namely \texttt{submit\_review} for the reviewer and \texttt{final\_tool} for the decider, whose docstrings are shown below.

\begin{docstring}[\texttt{submit\_review} (reviewer critic)]
Tool to submit the review of a single run (analysis agent + main agent exploration).
    Args:
        review: The complete, self-contained review of the run, following the requested structure.
    Returns:
        saved_result (dict): A message indicating the result of the save operation.
\end{docstring}

\begin{docstring}[\texttt{final\_tool} (decider critic)]
Tool to save the best run (analysis agent + main agent exploration).
    Args:
        best_run: The index of the best run, as numbered in the message. Starting from 1.
    Returns:
        saved_result (dict): A message indicating the result of the save operation.
\end{docstring}

\end{document}